\documentclass[aps,prb,twocolumn,floatfix]{revtex4}

\usepackage{amsmath}
\usepackage{amssymb}
\usepackage{graphicx}
\usepackage[squaren]{SIunits}

\providecommand{\vect}[1]{\ensuremath{\boldsymbol{#1}}}
\providecommand{\pdagger}{\phantom{\dagger}}

\providecommand{\mcC}{\ensuremath{\mathcal{C}}}
\providecommand{\mcD}{\ensuremath{\mathcal{D}}}
\providecommand{\mcE}{\ensuremath{\mathcal{E}}}
\providecommand{\mcF}{\ensuremath{\mathcal{F}}}
\providecommand{\mcG}{\ensuremath{\mathcal{G}}}
\providecommand{\mcH}{\ensuremath{\mathcal{H}}}

\providecommand{\mcO}{\ensuremath{\mathcal{O}}}
\providecommand{\mcS}{\ensuremath{\mathcal{S}}}
\providecommand{\mcU}{\ensuremath{\mathcal{U}}}

\DeclareMathOperator{\diag}{diag}
\DeclareMathOperator{\IM}{Im}
\DeclareMathOperator{\RE}{Re}
\DeclareMathOperator{\trace}{tr}

\usepackage[usenames,dvipsnames]{xcolor}

\begin{document}

\title{%
Interaction effects on almost flat surface bands in topological insulators
}

\author{Matthias Sitte}
\author{Achim Rosch}
\author{Lars Fritz}
\affiliation{%
Institut f{\"u}r Theoretische Physik,
Universit{\"a}t zu K{\"o}ln,
Z{\"u}lpicher Stra{\ss}e 77,
50937 K{\"o}ln, Germany
}

\date{November 11 2013}

\begin{abstract}
We consider ferromagnetic instabilities of two-dimensional helical Dirac fermions hosted on the surface of three-dimensional topological insulators.
We investigate ways to increase the role of interactions by means of modifying the bulk properties which in turn changes both the surface Dirac theory and the screening of interactions.
We discuss both the long-ranged part of the Coulomb interactions controlled by the dimensionless coupling constant $\alpha = e^{2}/(\hbar \epsilon v_{F}^{\mathrm{surf}})$ as well as the effects of local interactions parametrized by the ratio $U_{\mathrm{surf}}/D_{\mathrm{surf}}$ of a local interaction on the surface, $U_{\mathrm{surf}}$, and the surface bandwidth, $D_{\mathrm{surf}}$.
If large compared to $1$, both mechanisms can induce spontaneously surface ferromagnetism, thereby gapping the surface Dirac metal and inducing an anomalous quantum Hall effect.
We investigate two mechanisms which can naturally lead to small Fermi velocities $v_{F}^{\mathrm{surf}}$ and a corresponding small bandwidth $D_{\mathrm{surf}}$ at the surface when the bulk band gap is reduced.
The same mechanisms can, however, also lead to an enhanced screening of surface interactions.
While in all considered cases the long-ranged part of the Coulomb interaction is screened efficiently, $\alpha \lesssim 1$, we discuss situations, where $U_{\mathrm{surf}}/D_{\mathrm{surf}}$ becomes \emph{parametrically}\ large compared to $1$, thus inducing surface magnetism.
\end{abstract}


\maketitle

\section{Introduction}

Topological insulators (TIs) have attracted a great deal of attention in recent years as a topologically protected quantum state of matter originating from straightforward Bloch band theory.
Soon after the prediction of the two-dimensional (2D) quantum spin Hall insulator,\cite{Kane2005a, Kane2005b, Bernevig2006} the 2D TI was realized in HgTe quantum well heterostructures by K\"onig \textit{et al.}\cite{Koenig2007}
Characteristically, this quantum state of matter shows topologically protected spin currents running along the sample edges without net current flow.
Consequently, three-dimensional (3D) varieties of this new form of insulator were predicted\cite{Fu2007, Moore2007, Roy2009} and again realized soon afterwards in Bi$_{2}$Se$_{3}$ (Ref.~\citenum{Hasan2009}) and other compounds with strong spin-orbit coupling.\cite{Hasan2010, Qi2011}

The defining property of strong 3D TIs is the existence of a surface metal protected by the topology of the bulk bands in combination with time-reversal symmetry (TRS).
The surface is characterized by an odd number of band crossings which can be described by a 2D Dirac equation.
The surface metal shows a locking of momenta and spin degrees of freedom due to TRS, leading to a helical Dirac metal with vanishing backscattering and, most importantly, absence of electron localization.
It is therefore robust with respect to surface disorder of arbitrary strength as long as TRS is intact.\cite{Fritz2012}

Alternatively, 3D TIs can be characterized by the so-called $\theta$ angle which takes the value $\theta = \pi \bmod 2\pi$ for a strong TI, while $\theta = 0 \bmod 2\pi$ for a weak TI or an ordinary band insulator.
At the interface of a strong TI with vacuum, axion electrodynamics predicts a quantized anomalous Hall effect as a consequence of the varying $\theta$ term if the surface metal is gapped.\cite{Wilczek1987, Qi2008, Essin2009}
This surface state has many interesting properties: for example, when an electric charge approaches such a surface, it creates a magnetic field which can be described by a mirror charge of a magnetic monopole that has instead of the electric charge a quantized magnetic charge.\cite{Qi2008}
Due to the topological protection, a necessary condition to create the gapped surface state is to break TRS.
While this can be achieved by an external magnetic field, this has the disadvantage that also TRS is broken in the bulk which spoils the topological protection at least in the absence of inversion symmetry (IS).\cite{Sitte2012}
Signatures of a quantized Hall effect of such surfaces have been reported in HgTe which lacks IS.\cite{Bruene2011}
In Ref.~[\onlinecite{Sitte2012}], we have shown that the Hall response measured by contacting the edges remains quantized in such a system while, \textit{e.g.}, the monopole charge loses its quantization due to the breaking of TRS in the bulk.

It is therefore desirable to create TIs, where TRS is only broken on the surface.
For example, it has been suggested\cite{Qi2008} to coat the surface with a ferromagnetic insulating layer which is technologically demanding.
An alternative route is to search for TIs, where TRS is broken spontaneously on the surface, but not in the bulk, due to electron-electron interactions.
Indeed, it was shown that short-ranged electron-electron interactions may lead to a magnetic instability once a critical interaction strength is reached.\cite{Baum2012, Marchand2012, Schmidt2012}
Within this paper, we will also follow the latter route.

\begin{figure*}[tb]
\centering
\includegraphics[width=0.9\textwidth]{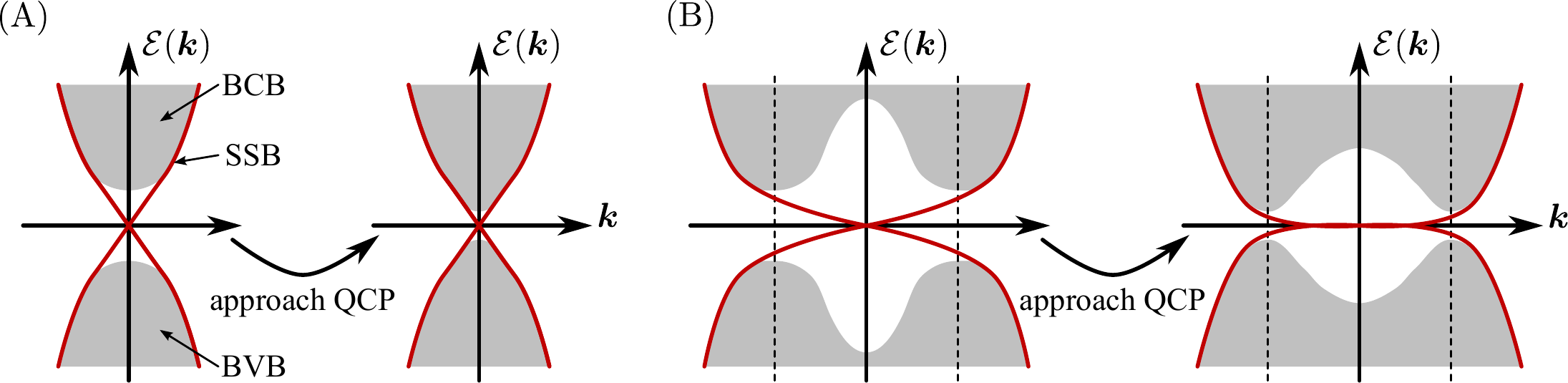}
\caption{%
(Color online)
Schematic illustration of the idea underlying a possible tuning of the surface Fermi velocity $v_F^{\mathrm{surf}}$ of 3D TIs:
As a function of some tuning parameter $m$ (\textit{e.g.}, spin-orbit coupling), we drive the TI towards the bulk quantum critical point (QCP), where it is converted into an ordinary band insulator.
At the critical point, the continuum of bulk conduction bands (BCB) and bulk valence bands (BVB) touch so that the surface state bands (SSB, red color) can ``unwind.''
The dashed vertical lines indicate the momenta, where the BCB and BVB touch.
We investigate under which circumstances close to the critical point the SSB can become flat, thereby potentially increasing the relative strength of interactions.
Importantly, in 3D TIs two classes of quantum phase transitions from the TI to the band insulator can be distinguished based on the (A) presence or (B) absence of bulk IS (Ref.~\citenum{Murakami2007}).
}
\label{fig:Idea}
\end{figure*}

The basic idea is to increase the effect of interactions by reducing the Fermi velocity of the surface bands.
In particular, we investigate the fate of the surface bands when the bulk band gap, $\mcE_{\mathrm{gap}}$, is reduced.
Two qualitatively different scenarios are shown in Fig.~\ref{fig:Idea}: in cases, where the surface Fermi velocity remains large at the bulk quantum critical point (QCP) $\mathcal{E}_{\mathrm{gap}} = 0$ [see Fig.~\ref{fig:Idea}~(A)], one can in general not expect surface magnetism.
A more interesting situation is depicted in Fig.~\ref{fig:Idea}~(B), where the surface bands get increasingly flat upon approaching the bulk QCP.
We will therefore investigate the following questions:
Under which conditions can one obtain small Fermi velocities of surface bands when the bulk band gap is reduced?
How does screening develop in this limit?
And, finally, does the interplay of flat bands and screening lead to magnetism on the surface?
We will mainly study two model systems with flat surface bands and consider both the role of long-ranged Coulomb and local interactions.

The interplay of flat bands and strong interactions can lead to many different states of matter: most famously, variants of the fractional quantum Hall effect can be induced in flat bands with non-zero Chern number.\cite{Regnault2011, Tang2011, Sun2011, Neupert2011, Venderbos2011, Parameswaran2013}
Our study focuses on a much simpler effect: the opening of a band gap at the Dirac point due to ferromagnetism.

Our main results are as follows:
(1)~In the most general case, surface bands do \emph{not}\ become flat close to the bulk QCP.
We identify, however, two cases where Fermi velocities do become small:
First, in the absence of IS one naturally obtains situations, where the bulk band structure and level repulsion of bulk and surface bands leads to small Fermi velocities.
This is realized in model~I motivated by the properties of strained 3D HgTe (Ref.~\citenum{Sitte2012}).
Second, in the presence of approximate chiral symmetries, almost flat surface bands are obtained for weak spin-orbit coupling.
This mechanism is discussed for the Fu-Kane-Mele model on a diamond lattice (model~II).
(2)~Concerning the effect of long-range interactions, the instability is controlled by the system-specific ``fine-structure constant'' $\alpha = e^{2}/(\hbar \epsilon v_{F}^{\mathrm{surf}})$ ($v_F^{\mathrm{surf}}$ denotes the surface Fermi velocity) which has to become large.
While we can decrease $v_{F}^{\mathrm{surf}}$ considerably, we find that we can not do so without rendering $\epsilon$ large.
In model~II, $\alpha$ remains constant in the limit $\mcE_{\mathrm{gap}} \to 0$ with a value depending on microscopic details, but vanishes in model~I.
As a consequence, we do not expect that the surface metal becomes critical and undergoes a surface instability due to long-range interactions.
(3)~However, even when the long-ranged part of the Coulomb interaction is well screened, the remaining short-range interactions can induce magnetism.
Here, we find that it is important whether the surface states remain confined to the boundaries of the system for $\mcE_{\mathrm{gap}} \to 0$.
In model~II, the penetration depth of the surface state remains finite in the limit, where the Fermi velocity at the surface vanishes, and thus a phase transition is induced at the surface.
In contrast, the penetration depth in the model~I diverges and surface magnetism is not enforced in general.

The organization of the paper is as follows:
We start with a discussion of the generic idea underlying our study in Sec.~\ref{sec:Idea}, and review the mechanism of spontaneous ferromagnetism due to long-range interactions, a phenomenon related to chiral symmetry breaking in QED, in Sec.~\ref{sec:CoulombInteractions}.
The same mechanism due to local Coulomb repulsion is discussed in in Sec.~\ref{sec:HubbardInteractions}.
In Sec.~\ref{sec:HgTeModel}, we turn to a concrete example for a system with flat surface bands motivated by strained 3D HgTe, discussing the effects of both long-range and short-range interactions.
Afterwards, in Sec.~\ref{sec:DiamondModel}, we discuss the Fu-Kane-Mele model on the 3D diamond lattice as a system with flat bands which originate from the presence of an additional symmetry at the critical point.
We conclude our discussions in Sec.~\ref{sec:Conclusions}.

\section{Ferromagnetic instability and tuning the interaction strength}
\label{sec:Idea}

We first discuss the possible mechanism of spontaneous ferromagnetism on the surface of TIs, driven by long-range Coulomb interactions (see Sec.~\ref{sec:CoulombInteractions}) or short-range Hubbard-type interactions (Sec.~\ref{sec:HubbardInteractions}).
The general idea we are pursuing can best be understood from a minimal model for the surface metal of 3D TIs which is a 2D Dirac theory for a single cone:
\begin{equation}
\label{eq:SurfaceHamiltonian}
H = v_{F}^{\mathrm{surf}} \sum_{\vect{k}} \Psi^{\dagger}_{\vect{k}} (k_{x} \sigma_{x} + k_{y} \sigma_{y}) \Psi^{\pdagger}_{\vect{k}},
\end{equation}
where $\Psi_{\vect{k}}$ is the two-component Dirac spinor, $v_{F}^{\mathrm{surf}}$ is the surface Fermi velocity, while $\sigma_{\alpha}$ ($\alpha \in \{ x,y,z \}$) denotes the usual Pauli matrices.
The effective surface Hamiltonian~(\ref{eq:SurfaceHamiltonian}) is gapless unless a term of the form
\begin{equation}
H' = r \sum_{\vect{k}} \Psi^{\dagger}_{\vect{k}} \sigma_{z} \Psi^{\pdagger}_{\vect{k}}
\end{equation}
is present which immediately opens a gap of size $2|r|$.
Such a mass term can be generated spontaneously by interactions corresponding to an order parameter $\langle \Psi^{\dagger} \sigma_{z} \Psi \rangle  \not=  0$.
The microscopic symmetry associated with it depends on the physical meaning of the entries of the Dirac spinors $\Psi_{\vect{k}}$.
On the surfaces of TIs, the spinor degrees of freedom are associated with the real spin due to the helical nature of the surface Dirac fermions, implying that order corresponds to spontaneous ferromagnetism which breaks TRS and opens a gap on the surface.
Once this gap is opened we expect to see, \textit{e.g.}, the surface magneto-electric effects associated with TIs, implying that there is an anomalous surface Hall conductivity given by $\sigma_{xy} = e^{2}/(2h)$.\cite{Qi2008, Essin2009}.
Note that in graphene the two components denote the sublattice degree of freedom, and thus order corresponds to a breaking of the sublattice IS.
Often, the term ``chiral symmetry breaking'' is used to describe such phase transitions in models described by Dirac equations.
In the following, we discuss two interaction-driven mechanisms to generate this kind of term spontaneously.

\subsection{Long-range Coulomb interactions}
\label{sec:CoulombInteractions}

Two-dimensional Dirac fermions interacting via long-ranged Coulomb interactions,
\begin{equation}
\begin{split}
H &= \int \frac{d^2 k}{(2\pi)^2} \, \Psi^{\dagger}_{\vect{k}} (\vect{k} \cdot \vect{\sigma}) \Psi^{\pdagger}_{\vect{k}}
+ \frac{\alpha}{4\pi} \int d^2 x \, d^2 x' \, \frac{\rho(\vect{x}) \rho(\vect{x}')}{|\vect{x} - \vect{x}'|}
\end{split}
\end{equation}
with the fermion density operator $\rho(\vect{x}) = \psi^{\dagger}(\vect{x})\psi(\vect{x})$, are unstable towards the breaking of chiral symmetry once the dimensionless effective fine-structure constant controlling the relative strength of Coulomb interactions,
\begin{equation}
\label{eq:FineStructureConstant}
\alpha = \frac{e^{2}}{\hbar \epsilon_{0} \epsilon v_{F}^{\mathrm{surf}}},
\end{equation}
exceeds a critical value $\alpha_{c}$ (Refs.~\citenum{Khveshchenko2001, Son2007, Drut2009}).
In the above expression, $e$ is the electron charge, $\epsilon$ the mean dielectric constant of the surrounding media, and $v_{F}^{\mathrm{surf}}$ denotes the Fermi velocity of the surface Dirac fermions.
The critical interaction strength $\alpha_{c}$ for chiral symmetry breaking remains finite as long as the number of fermion flavors does not exceed a critical number $N_{c}$, where for $N > N_{c}$ the semimetal is always stable and long-range interactions cannot trigger an instability.
For graphene ($v_F \approx \unit{10^6}{\meter\per\second}$, $N = 4$) this transition has been discussed intensively with the theoretical prediction for $\alpha_{c} \approx 0.8-3.5$, depending on the approximation scheme\cite{Wang2012b}.

Concerning Eq.~\eqref{eq:FineStructureConstant}, we observe that there are in principle two strategies of increasing the fine-structure constant $\alpha$ by either decreasing the dielectric constant $\epsilon$ and/or reducing the surface Fermi velocity $v_{F}^{\mathrm{surf}}$.
The first strategy has been pursued in the context of graphene: suspending the sample modifies the dielectric environment, and one expects a value of $\alpha \approx 2.2$ which, in principle, is well within the range of possible values for $\alpha_{c}$, but current experimental evidence does not support the existence of an excitonic insulator.
For example, recent quantum oscillation measurements show semimetallic behavior down to lowest fillings and temperatures.\cite{Elias2011}

Our basic idea is to pursue the second strategy, \textit{i.e.}, we try to decrease $v_{F}^{\mathrm{surf}}$, hoping that the dielectric constant $\epsilon$ of the bulk material remains largely constant.
However, we find that the latter is a strong restriction and usually can not be achieved in a generic setting.
We show that, under certain circumstances, one can tune the Fermi velocity $v_{F}^{\mathrm{surf}}$ by means of tuning the TI towards its bulk QCP, where it is converted into an ordinary band insulator.
The appropriate knob is spin-orbit coupling which has been demonstrated via chemical substitution in bismuth-based compounds,\cite{Xu2011} but can in principle also be done in HgTe systems, where Hg is replaced by Cd (Ref.~\citenum{Koenig2007}).

\subsection{Short-range Hubbard-type interactions}
\label{sec:HubbardInteractions}

Similarly, one can discuss the effect of short-ranged interactions, where we consider the following model:
\begin{equation}
H = v_{F}^{\mathrm{surf}} \int \frac{d^{2} k}{(2\pi)^{2}} \, \Psi^{\dagger}_{\vect{k}} (\vect{k} \cdot \vect{\sigma}) \Psi^{\pdagger}_{\vect{k}} + U \int d^{2} x\ \rho(\vect{x}) \rho(\vect{x}).
\end{equation}
In that situation, one can use a simple mean-field theory to obtain a crude estimate for the critical value of the interaction constant $U_{c} \propto v_{F}^{\mathrm{surf}}$ (Ref.~\citenum{Herbut2006}).
Starting from the Dyson-Schwinger equation for a dynamic mass term $r$ for a purely local interaction, one obtains the following mean field equation:
\begin{multline}
r(\epsilon, \vect{p}) = i \int_{-\infty}^{\infty} \frac{d\nu}{2\pi} \int \frac{d^{2}k}{(2\pi)^{2}} \\
\frac{U\ r(\nu, \vect{k})}{\nu^{2} - (v_{F}^{\mathrm{surf}} |\vect{k}|)^{2} - r(\nu, \vect{k})^{2} + i0^{+}}.
\end{multline}
This gives a Stoner-type criterion for the relation between critical interaction strength $U_{c}$ and the Fermi velocity $v_{F}^{\mathrm{surf}}$ reading as
\begin{equation}
1 = \frac{U_{c}}{4\pi} \frac{\Lambda}{v_{F}^{\mathrm{surf}}}
\end{equation}
with $\Lambda$ being the ultraviolet momentum cutoff.
This implies that upon reducing the velocity $v_{F}^{\mathrm{surf}}$ of the surface Dirac fermions, the critical interaction strength $U_{c}$ is lowered, and thus this seems to be the promising way towards driving a ferromagnetic surface instability.
A phase transition at the surface of a TI is therefore expected when the local interactions on the surface $U_{\mathrm{surf}}$ become larger than the surface bandwidth $D_{\mathrm{surf}} \sim v_F^{\mathrm{surf}} \Lambda$.
As we will discuss below, it is, however, important to realize that the local surface interactions can become much weaker than the bulk interactions when the surface state wave function penetrates several layers into the bulk.

Overall, it appears therefore desirable to reduce the Fermi velocity of the surface electron as much as possible to drive the quantum phase transition towards a gapped surface state.

\section{Model~I: Minimal model for strained 3D H\lowercase{g}T\lowercase{e}}
\label{sec:HgTeModel}

We first discuss a TI, whose effective model is inspired by the symmetries of strained 3D HgTe.
The topological surface states of strained 3D HgTe can be described in terms of the $\Gamma_{6}$ electron ($E$) bands and the $\Gamma_{8}$ light-hole ($LH$) bands.
We neglect the coupling of the $E$ bands to the $\Gamma_{8}$ heavy-hole ($HH$) bands, because their presence only changes the quantitative features of the band structure, but does not alter the presence of the surface states as long as the strain-induced gap is open.\cite{Dai2008}
Therefore, limiting ourselves to the following four-dimensional set of basis states (for details see Ref.~\onlinecite{Sitte2012}),
\begin{equation}
\begin{split}
|{1}\rangle &\equiv |{\Gamma_{6}, +\tfrac{1}{2}}\rangle = |{s, \uparrow}\rangle, \\
|{2}\rangle &\equiv |{\Gamma_{6}, -\tfrac{1}{2}}\rangle = |{s, \downarrow}\rangle, \\
|{3}\rangle &\equiv |{\Gamma_{8}, +\tfrac{1}{2}}\rangle = (|{p_{x}, \downarrow}\rangle + i |{p_{y}, \downarrow}\rangle - 2 |{p_{z}, \uparrow}\rangle)/\sqrt{6}, \\
|{4}\rangle &\equiv |{\Gamma_{8}, -\tfrac{1}{2}}\rangle = -(|{p_{x}, \uparrow}\rangle - i |{p_{y}, \uparrow}\rangle + 2 |{p_{z}, \downarrow}\rangle)/\sqrt{6},
\end{split}
\end{equation}
the $4 \times 4$ Bloch Hamiltonian of the effective model can conveniently be expressed in terms of the identity, $\openone$, five Dirac matrices $\Gamma_{a}$ satisfying the Clifford algebra $\{ \Gamma_{a}, \Gamma_{b} \} = 2 \delta_{a b}\ \openone$, and their ten commutators $\Gamma_{ab} \equiv [\Gamma_{a}, \Gamma_{b}]/(2i)$.
Using
\begin{equation}
\Gamma^{(0,1,2,3,4)} \equiv (\tau_{z} \otimes \sigma_{0}, -\tau_{x} \otimes \sigma_{y}, \tau_{x} \otimes \sigma_{x}, \tau_{y} \otimes \sigma_{0}, \tau_{x} \otimes \sigma_{z}),
\end{equation}
where $\vect{\tau} = (\tau_{x}, \tau_{y}, \tau_{z})^{T}$ and $\vect{\sigma} = (\sigma_{x}, \sigma_{y}, \sigma_{z})^{T}$ act on the orbital and spin degrees of freedom, respectively, and $\tau_{0} = \sigma_{0} = \openone$ denotes the $2 \times 2$ identity matrix, one obtains the following minimal model:
\begin{align}
\label{eq:MinimalHgTeModel}
\nonumber
H &= -t \sum_{\vect{R}} \sum_{j=1}^{3} \biggl[ \Psi^{\dagger}_{\vect{R} + \hat{\vect{e}}_{j}} \biggl( \frac{\Gamma_{0} - i \Gamma_{j}}{2} \biggr) \Psi^{\pdagger}_{\vect{R}}  + \mathrm{H.c.} \biggr] \\
&\quad + \sum_{\vect{R}} \Psi^{\dagger}_{\vect{R}} (-\mu\ \openone + m\ \Gamma_{0} + \Delta\ \Gamma_{04}) \Psi^{\pdagger}_{\vect{R}}\;.
\end{align}
Here, $t$ is the overlap parameter describing hopping between nearest neighbors on a simple cubic lattice, and $m$ denotes the so-called tuning parameter.
$\Delta$ parametrizes the breaking of bulk IS, and a finite chemical potential $\mu$ breaks particle-hole symmetry.

\subsection{Closing of the bulk band gap}

Following the theoretical scheme of quantum phase transitions of 3D TIs by Murakami,\cite{Murakami2007} we can identify two types of criticality within model~I: (A) the bulk band gap closes at one of the time-reversal-invariant momenta (TRIMs), or (B) the bulk band gap closes away from any TRIM at several points in the Brillouin zone.
Whether a system belongs to either of the two classes depends on the presence (A) or absence (B) of bulk IS ($\Delta/t = 0$ and $\Delta/t \not= 0$, respectively, in model~I).
The general situation is shown in Fig.~\ref{fig:Idea}.
Interestingly, only in case~(A) one finds a direct transition from the TI to the trivial band insulator, while in case~(B) there is generically an intermediate semi-metallic phase described by Weyl fermions.
The reason for the intermediate metallic phase is that one can associate topological charges with three-dimensional Weyl points.
After touchdown these points start to move around in the Brillouin zone until they meet a partner with which they can annihilate.\cite{Murakami2007}
This generic scenario can be modified if further symmetries are present (see following).

In case~(A), exemplified by the bismuth-based 3D TIs, the surface Fermi velocity $v_{F}^{\mathrm{surf}}$ is fixed by the bulk Fermi velocity $v_{F}^{\mathrm{bulk}}$ and therefore remains finite at the QCP.
For model~I, Eq.~(\ref{eq:MinimalHgTeModel}), we observe that we are in case~(A) for $\Delta/t = 0$ with the model describing a trivial band insulator for $|m/t| > 3$, a strong TI for $1 < |m/t| < 3$, and a weak TI for $|m/t| < 1$.
The bulk band structure is characterized by a degenerate 3D Dirac point sitting at the one of the TRIMs $(0,0,0)^{T}$, $(\pi, 0, 0)^{T}$, $(\pi, \pi, 0)^{T}$ or $(\pi, \pi, \pi)^{T}$ for $m/t = 3, 1, -1, -3$, respectively [see Fig.~\ref{fig:Idea}~(A)].
The surface state obtained in the topological phase is located at the TRIM as well, and the projected surface theory simply inherits the Fermi velocity of the bulk theory, \textit{i.e.}, $v_{F}^{\mathrm{surf}} \sim v_{F}^{\mathrm{bulk}}$.

\begin{figure*}[tb]
\centering
\includegraphics[width=0.45\textwidth]{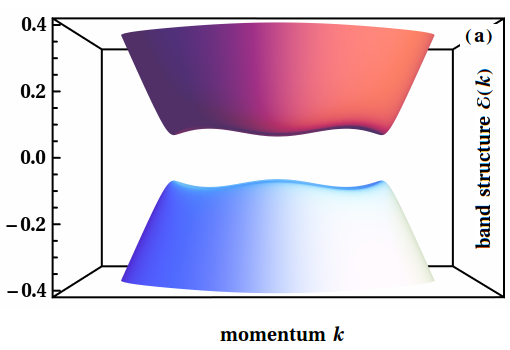}
\hspace*{0.05\textwidth}
\includegraphics[width=0.45\textwidth]{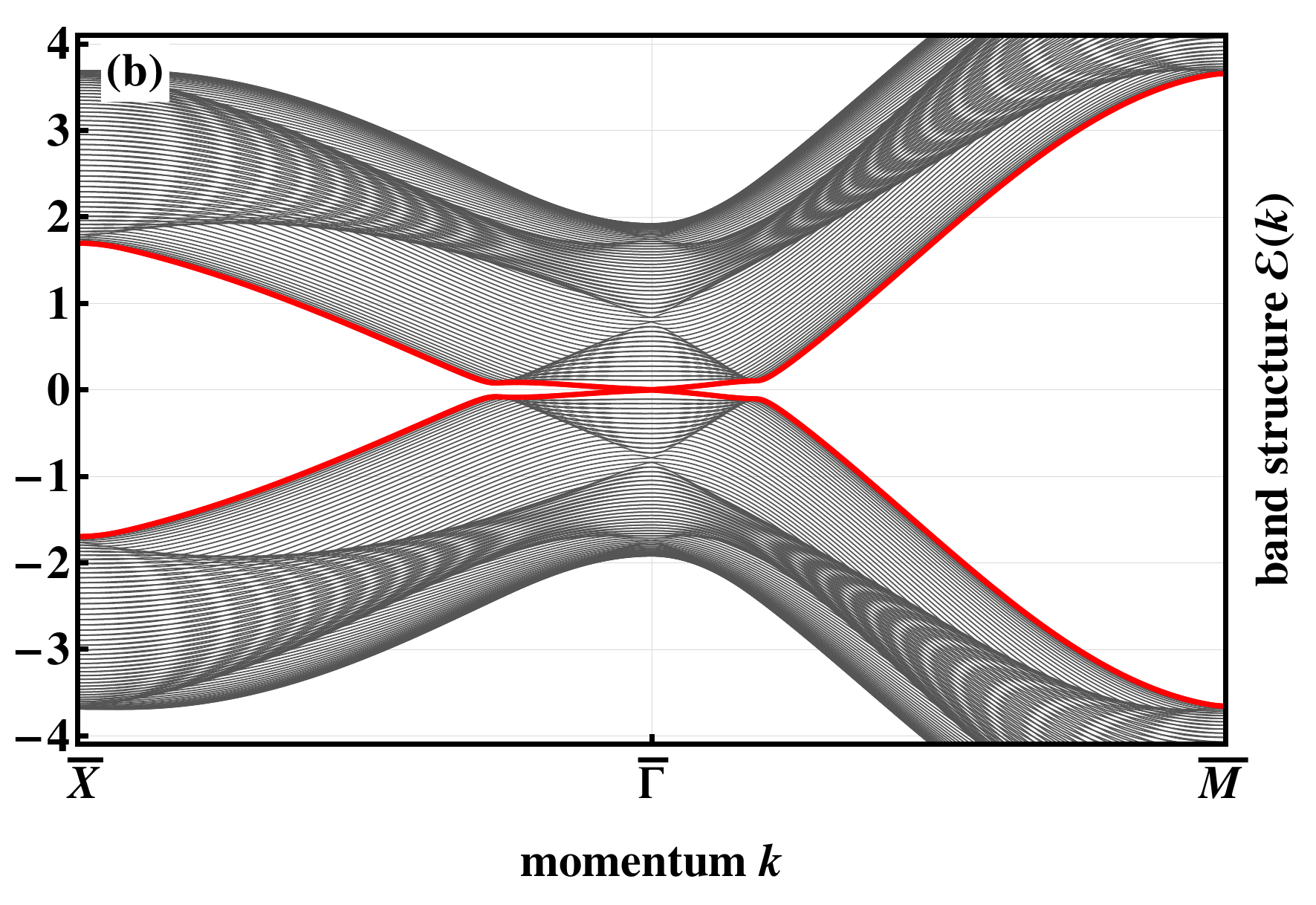}
\caption{%
(Color online)
\textbf{(a)}~Projection of the bulk band structure of model~I [Eq.~(\ref{eq:MinimalHgTeModel})] onto the $(001)$ surface Brillouin zone in the STI phase for $t = 1$, $m/t = 2.6$, $\Delta/t = 0.7$, and $\mu/t = 0$.
To leading order, the bulk band gap closes in a stamp-like way, where higher-order terms lead to a fine structure of the stamp on the order $(\Delta/t)^{2}$.
\textbf{(b)}~Slab band structure of the minimal model for a slab of $50$ unit cells in the $[001]$ direction in the STI phase.
The surface-state bands (indicated by red color) clearly traverse the bulk band gap and become flatter and flatter as we drive the system towards the bulk QCP, $m \to m_{c1}$.
}
\label{fig:MinimalHgTeModel}
\end{figure*}

In case~(B), however, the surface band connects two (massive) bulk Dirac cones separated by a finite momentum which opens room for small surface velocities.
In model~I, this case is obtained for $\Delta/t \not= 0$.
In momentum space, the Bloch Hamiltonian corresponding to the Hamiltonian~(\ref{eq:MinimalHgTeModel}) is given by
\begin{equation}
\mcH(\vect{k}) = \epsilon(\vect{k}) \openone + \sum_{j=0}^{4} d_{j}(\vect{k}) \Gamma_{j} + \Delta \, \Gamma_{04},
\end{equation}
with
\begin{equation}
\begin{gathered}
\epsilon(\vect{k}) = -\mu, \;
d_{0}(\vect{k}) = m - t (\cos k_x + \cos k_y + \cos k_z), \\
d_{1}(\vect{k}) = -t \sin k_x, \;
d_{2}(\vect{k}) = -t \sin k_y, \;
d_{3}(\vect{k}) = -t \sin k_z.
\end{gathered}
\end{equation}
Aside from a trivial shift of the energies by $\epsilon(\vect{k})$, the spectrum of the Hamiltonian consists of four non-degenerate eigenvalues $\mcE_{\pm\pm}(\vect{k})$ taking the values
\begin{multline}
\mcE_{\pm\pm}(\vect{k}) = \epsilon(\vect{k}) \pm \biggl[ d_{0}(\vect{k})^{2} \\
+ \Bigl( \Delta \pm \sqrt{ d_{1}(\vect{k})^{2} + d_{2}(\vect{k})^{2} + d_{3}(\vect{k})^{2}} \Bigr)^2 \biggr]^{1/2}.
\end{multline}
To close the band gap in the particle-hole symmetric case, \textit{i.e.}, for $\epsilon(\vect{k}) = 0$, one needs to fulfill the two conditions $d_{0}(\vect{k}) = 0$ and $\Delta - \sqrt{d_{1}(\vect{k})^{2} + d_{2}(\vect{k})^{2} + d_{3}(\vect{k})^{2}} = 0$ simultaneously.
Both equations individually define two-dimensional manifolds of solutions in momentum space, and for fixed values of $\Delta/t$ and $m/t$ we find three possible situations:
(i)~There is no point in momentum space fulfilling both conditions at a time, implying that we are either in the TI or in the trivial insulator phase.
(ii)~If we find lines of solutions in  momentum space, then we are in an intermediate metallic phase.
Note that for a general $\epsilon(\vect{k})$ those lines are again reduced to points, and both the critical point as well as the intermediate metallic phase are characterized by isolated Dirac points.
(iii)~We find isolated solutions to both equations describing a single Dirac point.
Then, we are at the critical point from the metallic phase to one of the two adjacent insulating phases.

Consequently, we can define three domains for the tuning parameter $m$ in the vicinity of the bulk phase transition:
$m < m_{c1}$ as TI, $m > m_{c2}$ as trivial insulator, and $m_{c1} < m < m_{c2}$ as the intermediate metallic phase characterized by a number (possibly a line) of Weyl points and a semimetallic density of states.

For the minimal HgTe model~(\ref{eq:MinimalHgTeModel}) we find that coming from the strong TI phase $m < m_{c1}$ the bulk band gap closes at
\begin{equation}
m_{c1}/t = 2 + \cos \arcsin (\Delta/t) = 2 + \sqrt{1 - (\Delta/t)^{2}}
\end{equation}
on the $\langle 100 \rangle$ lines at a distance $|\vect{Q}_1| = \arcsin(\Delta/t)$.
When further increasing the tuning parameter $m/t$ we move through the semimetallic phase until the bulk band gap re-opens at
\begin{align}
m_{c2}/t = 3 \cos \arcsin (\Delta/\sqrt{3}t) = 3 \sqrt{1 - (\Delta/t)^{2}/3}\;.
\end{align}
This happens on the $\langle 111 \rangle$ lines at a distance $|\vect{Q}_2| = \sqrt{3} \arcsin(\Delta/\sqrt{3}t)$.

\subsection{Flatness of surface bands}

Tuning the mass $m$ of the system towards $m_{c1}$ coming from the TI ($m < m_{c1}$), the bulk Dirac cones are gapped with a band gap of size $\mcE_{\mathrm{gap}} = 2 \sqrt{1 - (\Delta/t)^{2}} |m - m_{c1}|$.
In the limit of weak IS breaking, $\Delta/t \ll 1$, the bulk band gap closes on a sphere of radius $\Delta/v_{F}^{\mathrm{bulk}}$ in momentum space, where $v_{F}^{\mathrm{bulk}} = t a$.
Projected onto the surface, this implies an approximate constant bulk band gap for all surface momenta smaller than $\Delta/v_{F}^{\mathrm{bulk}}$.
In Fig.~\ref{fig:MinimalHgTeModel}~\textbf{(a)}, we show the projection of the bulk bands onto a surface Brillouin zone.
Even for the rather large value of $\Delta$ used here, $\Delta/t = 0.7$, one can see that the bulk band gap remains approximately constant as a function of the surface momentum up to small wiggles discussed in the following.
Due to level repulsion, the surface modes will have energies within the bulk band gap which enforces small velocities of the surface-state bands [see Fig.~\ref{fig:MinimalHgTeModel}~\textbf{(b)}].
One can visualize this the following way: the bulk bands act as two ``stamps'' in energy space which have a distance of $\mcE_{\mathrm{gap}}$, and the surface bands therefore have to become flatter and flatter when $\mcE_{\mathrm{gap}}$ shrinks.
The ``spectral pressure'' of the bulk bands therefore leads to a flattening of the surface bands.

\begin{figure*}[tb]
\centering
\includegraphics[width=0.45\textwidth]{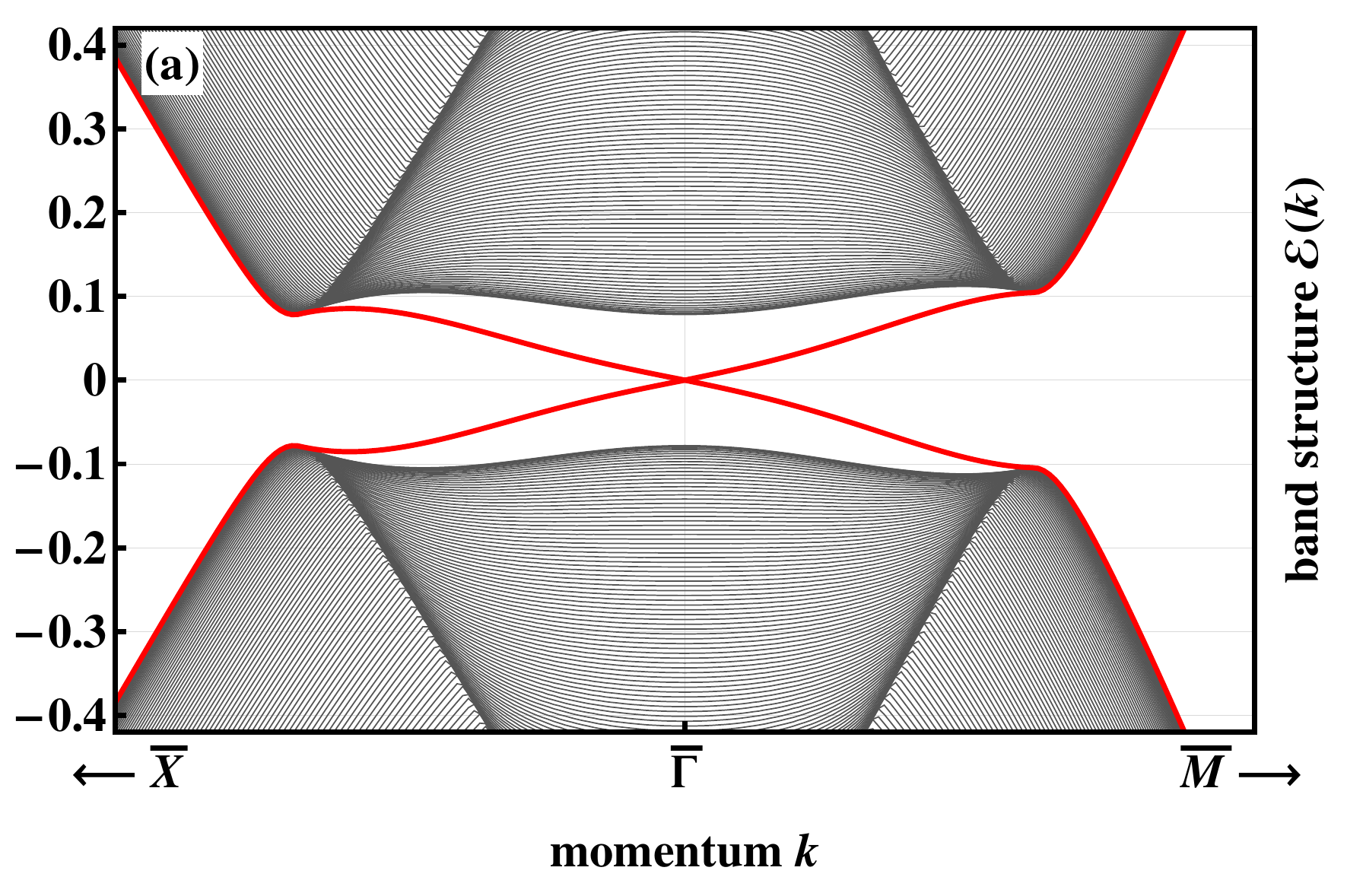}
\hspace*{0.05\textwidth}
\includegraphics[width=0.45\textwidth]{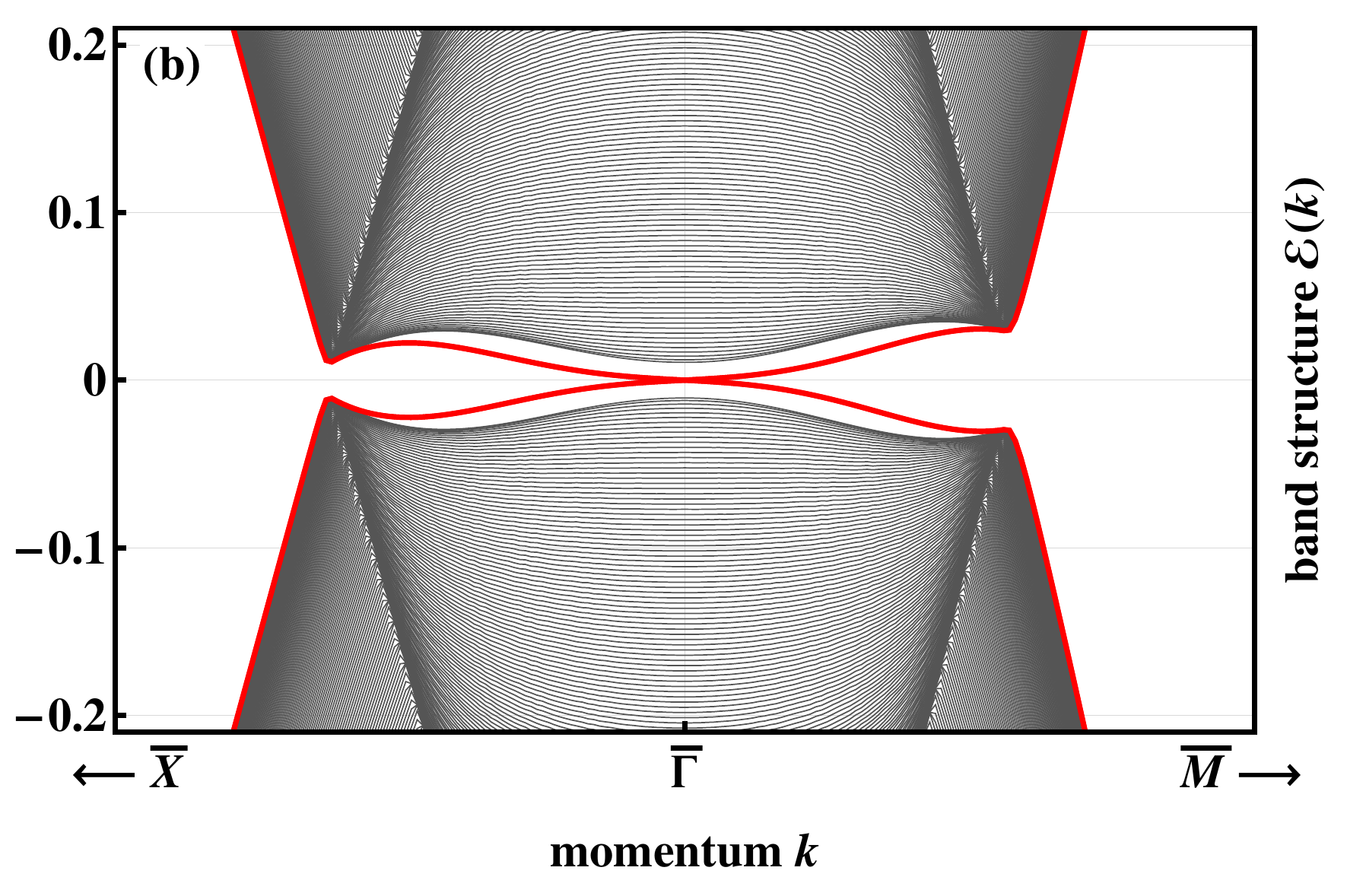}
\caption{%
(Color online)
Band structure of the minimal model~I [Eq.~(\ref{eq:MinimalHgTeModel})] for a slab of $1000$ unit cells in the $[001]$ direction in the STI phase for $t = 1$, $\Delta/t = 0.7$, $\mu/t = 0$ and \textbf{(a)} $m/t = 2.6$ and \textbf{(b)} $m/t=2.7$ (the critical value of the tuning parameter is $m_{c1}/t = 2.71414$).
The surface-state bands (red color) are confined to move within the narrow constriction of the bulk band gap, leading to small Fermi velocities of the order $v_{F}^{\mathrm{surf}} \lesssim v_{F}^{\mathrm{bulk}} |(\mcE_{\mathrm{gap}} + \mcE_{\mathrm{wiggle}})/\Delta| \sim v_{F}^{\mathrm{bulk}} |\Delta/t|^{3}$ for $\mcE_{\mathrm{gap}} \to 0$.
}
\label{fig:MinimalHgTeModel2}
\end{figure*}

An estimate for the surface Fermi velocity is obtained from a simple linear interpolation between the two edges of the projected bulk Dirac cones [see Fig.~\ref{fig:MinimalHgTeModel2}]:
\begin{equation}
v_{F}^{\mathrm{surf}} \lesssim v_{F}^{\mathrm{bulk}} \biggl| \frac{\mcE_{\mathrm{gap}} + \mcE_{\mathrm{wiggle}}}{\Delta} \biggr|,
\end{equation}
where in our model $v_{F}^{\mathrm{bulk}} = t a$ with lattice spacing $a$, and $\Delta$ is a constant parametrizing the bulk inversion asymmetry.
$\mcE_{\mathrm{wiggle}}$ is the size of the wiggles parameterizing the roughness of the stamp, \textit{i.e.}, the variation of the size of a bulk band gap as function of the surface momentum of the bulk band gap.
Due to these wiggles, the surface bands do not become exactly flat in the limit $\mcE_{\mathrm{gap}} \to 0$.
We find that $\mcE_{\mathrm{wiggle}} \propto \Delta^4$, and therefore we obtain for small, finite $\Delta$ in the limit $\mcE_{\mathrm{gap}} \to 0$ a surface velocity
\begin{equation}
v_{F}^{\mathrm{surf}} \sim v_{F}^{\mathrm{bulk}} |\Delta/t|^3.
\end{equation}
We have checked that the picture is not modified when a finite $\epsilon(\vect{k})$ breaking particle-hole symmetry is taken into account.

\subsection{Long-range Coulomb interactions and screening}

To estimate the strength of interactions for small $\mcE_{\mathrm{gap}}$, we have to investigate the role of screening by bulk states in this limit.
To develop an understanding of screening in 3D TIs, we have computed the dielectric constant $\epsilon(\omega, \vect{q})$ in the static limit ($\omega = 0$) in the limit $\mcE_{\mathrm{gap}} > \mcE_{\mathrm{wiggle}}$, where the bulk band gap is approximately independent of momentum on the surface of a sphere in momentum space.
Within the random phase approximation, the static polarization function $\Pi(\omega = 0, \vect{q})$ shows a divergence close to the critical point (see Appendix~\ref{app:PolarizationFunctionModelI}):
\begin{equation}
\Pi(\omega = 0, \vect{q}) \propto \frac{\vect{q}^{2}}{\mcE_{\mathrm{gap}}^{2}},
\end{equation}
implying that the bulk dielectric constant diverges for small $\mcE_{\mathrm{gap}} > \mcE_{\mathrm{wiggle}}$,
\begin{equation}\label{eps}
\epsilon(\vect{q}) = 1 + \frac{4 \pi e}{|\vect{q}|^{2}}\ \Pi(\omega = 0, \vect{q}) = c_{0} + c_{2}  \frac{t^2}{\mcE_{\mathrm{gap}}^{2}},
\end{equation}
where $c_{0,2}$ are numerical prefactors.
Consequently, the effective interaction strength $\alpha$ goes to zero instead of diverging as we approach the bulk critical point for small $\mcE_{\mathrm{gap}} > \mcE_{\mathrm{wiggle}}$:
\begin{equation}
\alpha = \frac{e^{2}}{\hbar \epsilon_{0} \epsilon v_{F}^{\mathrm{surf}}} \sim \frac{|\Delta| \mcE_{\mathrm{gap}}}{t^2} \ll 1.
\end{equation}
In the other limit $\mcE_{\mathrm{gap}} < \mcE_{\mathrm{wiggle}}$, both the surface velocity and the dielectric constant $\epsilon$ become independent of $\mcE_{\mathrm{gap}}$ due to the vanishing density of states at the Dirac points.
Therefore, $\alpha$ remains much smaller than $1$ also in this limit.

In this section, we have shown that weak IS breaking naturally leads to flat surface bands by spectral pressure of the bulk bands when the bulk band gap becomes small.
Nevertheless, efficient screening renders the long-ranged part of the Coulomb interactions weak.

\subsection{Effect of local interactions}

As long-ranged interactions are effectively screened in model~I, one has to consider short-ranged interaction.
As argued in Sec.~\ref{sec:HubbardInteractions}, a surface phase transition can be expected when the local surface interactions, $U_{\mathrm{surf}}$, become larger than the effective bandwidth of the surface states, $D_{\mathrm{surf}}$.
The latter becomes small, $D_{\mathrm{surf}} \sim \mcE_{\mathrm{gap}}$, in the limit of small $\mcE_{\mathrm{gap}} > \mcE_{\mathrm{wiggle}}$.
For an estimate of $U_{\mathrm{surf}}$ one has to take into account that the wave function penetrates deeply into the bulk in this limit.
For a given envelope wave function $\Psi_{E}(z)$ describing the penetration of the surface state in the $z$ direction into the bulk, a crude estimate of the effective surface interaction is obtained from $U_{\mathrm{surf}} \approx \int |\Psi_{E}(z)|^{4}\ U_{\mathrm{bulk}}\ dz \sim  U_{\mathrm{bulk}} (a/d)$, where $d$ is the penetration depth and $a$ the lattice spacing.
Here, we used that $\Psi_{E}(z) \sim 1/\sqrt{d}$ as $\int |\Psi_{E}(z)|^{2}\ dz = 1$.
The penetration depth will diverge for small $\mcE_{\mathrm{gap}}$, \textit{i.e.}, $d \sim v_{F}^{\mathrm{bulk}}/\mcE_{\mathrm{gap}}$, and therefore both $U_{\mathrm{surf}}$ and $D_{\mathrm{surf}}$ are expected to vanish linearly for small $\mcE_{\mathrm{gap}}$.

This simple hand-waving argument suggests that also local interactions will generically not drive surface instabilities in model~I.
Our estimate has, however, completely ignored the coupling of bulk and surface modes and the role of higher momentum states.
These effects can, however, be taken into account within a straightforward random-phase approximation (RPA) calculation for a TI slab which includes automatically the coupling of surface and bulk modes by interactions.
As usual, for local interactions the (non-self-consistent) RPA calculation gives for the critical interaction the same results as a straightforward mean field calculation.

For simplicity, we model the interactions in the bulk by a local, intra-orbital Hubbard interaction
\begin{equation}
H_{\mathrm{int}} = U \sum_{i, \alpha} \hat{n}_{i \alpha \uparrow}^{\pdagger} \hat{n}_{i \alpha \downarrow}^{\pdagger},
\end{equation}
where $\alpha \in \{ E, LH \}$ denotes the orbital degree of freedom.
To determine the leading instability in model~I within RPA (or, equivalently, within a mean field approach), we have computed for a given surface momentum $\vect q$ the layer-resolved susceptibility matrix $\chi$ corresponding to the order parameter $\langle \Psi^{\dagger} \sigma_{z} \Psi \rangle$ in a slab geometry:
\begin{equation}
\label{eq:SusceptibilityMatrix}
\chi_{j\pm,j' \pm}(\vect{q}) = \int \frac{d^{2}k}{(2\pi)^{2}} \!\!\! \sum_{\mcE_{\vect{k},\alpha} < 0 < \mcE_{\vect{k} + \vect{q}, \beta}} \!\!\!\!\!\!\!\! \frac{M^{\alpha \beta}_{\vect k,\vect k+\vect q}(j \pm)M^{\beta \alpha}_{\vect k+\vect q,\vect k}(j' \pm)}{\mcE_{\vect{k}, \alpha} - \mcE_{\vect{k} + \vect{q}, \beta}}.
\end{equation}
Here, $M^{\alpha \beta}_{\vect{k}, \vect{k}'}(j \pm) = \langle{\vect{k} \alpha}|  \mcS_{\pm}(j) |{\vect{k}' \beta}\rangle $ denotes the matrix elements of the spin operator in layer $j$ with respect to the $|{\vect{k}, \alpha}\rangle$ eigenstates of the Hamiltonian for a given momentum $\vect k$ of the surface Brillouin zone.
For the $E$ and $LH$ orbitals, respectively, the operator $\mcS_{\pm}$ is defined as
\begin{equation}
\mcS_{\pm} \equiv \biggl( \frac{\tau_{0} \pm \tau_{z}}{2} \biggr) \otimes \sigma_{z}.
\end{equation}
The following calculations have been performed for $t = 1$ and $\Delta/t = 0.7$ for concreteness, but the qualitative picture also holds for other parameter sets.

Within RPA, a phase transition takes place when the largest eigenvalue of $\chi(\vect q)$ becomes larger than $1/U$.
We find that all eigenvalues of $\chi$ are of the same order, and most importantly none of them diverges close to the critical point.
The two eigenmodes of the largest, doubly degenerate eigenvalue are localized on the surfaces of the slab and have the in-plane momentum $\vect{q}_{\mathrm{surf}} = (0,0)^{T}$ describing a ferromagnetic instability on the surface(s).
The eigenmode of the largest bulk eigenvalue on the other hand shows a rapidly oscillating behavior as s function of transverse layer index, indicating an anti-ferromagnetic component with momentum $\vect{q}_{\mathrm{bulk}} = (0,0,\pi)^{T}$.
We have confirmed that this instability can also be obtained from a bulk mean-field calculation.

\begin{figure}[tb]
\centering
\includegraphics[width=0.45\textwidth]{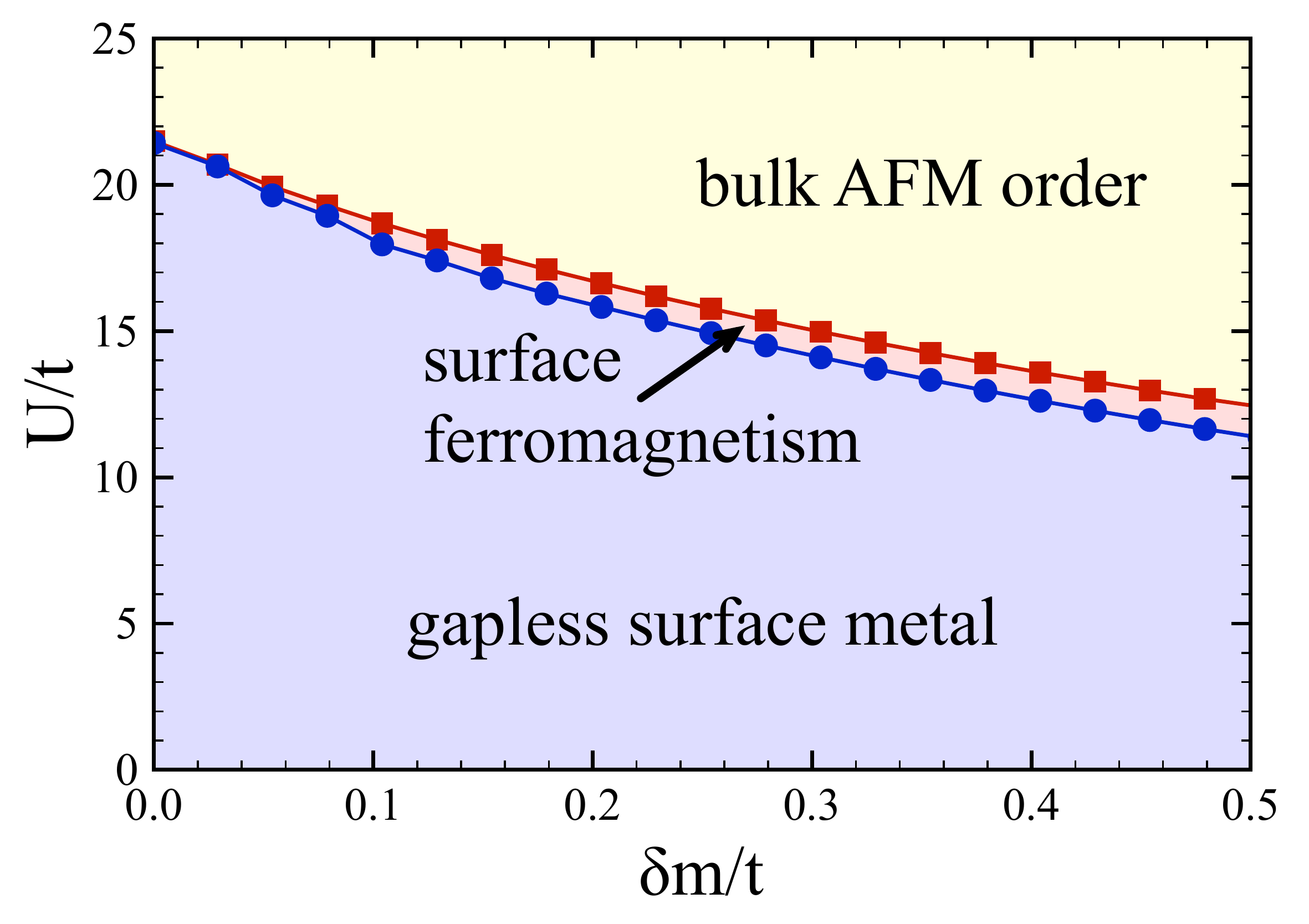}
\caption{%
(Color online)
Qualitative phase diagram for model~I computed from the layer-resolved susceptibility matrix~(\ref{eq:SusceptibilityMatrix}) for a slab of $N_{z} = 40$ layers with parameters $t = 1$ and $\Delta/t = 0.7$.
The leading bulk instability is  anti-ferromagnetism at $\vect q=(0,0,\pi)$.
Only in a tiny parameter regime surface ferromagnetism occurs despite the fact that the surface Fermi velocity becomes tiny for small $\delta m/t$ (see Fig.~\ref{fig:MinimalHgTeModel2}).
In the vicinity of the critical point the surface modes penetrate deeper and deeper into the bulk of the system so that the leading instability of the surface modes (red squares) and of the bulk modes (blue circles) merge.
}
\label{fig:SusceptibilityMatrix}
\end{figure}

The resulting phase diagram is shown in Fig.~\ref{fig:SusceptibilityMatrix} as a function of interaction strength $U$ and distance $\delta m$ from the point, where the bulk gap closes.
For the specific model considered here, we find that the leading instability of the system is surface ferromagnetism which occurs as a precursor of bulk anti-ferromagnetism.
There is, however, only a tiny range of  parameters, where the surface is magnetic and the bulk is not.
Consistent with our analytic estimates we find, however, no strong enhancement of surface magnetism despite the fact that the surface state bands are almost completely flat for small $\delta m$.

As a similar interaction strength induces a bulk or a surface transition, the predictive power of our calculation is limited: changes in the surface chemistry, modification of band structure, changes of inter and intra-orbital interactions, and corrections beyond mean field can change the overall phase diagram.
This will be different in the second model considered below.

\section{Model~II: Fu-Kane-Mele model on the diamond lattice}
\label{sec:DiamondModel}

In this section, we want to investigate an alternative mechanism which can lead to flat surface bands.
In two dimensions, exactly flat surface bands are well known in particle-hole symmetric models for graphene when only nearest-neighbor hopping is taken into account.\cite{Peres2006}
Similarly, generalizations of graphene as quantum spin Hall insulator to higher-dimensional lattices, \textit{e.g.}, the diamond lattice\cite{Fu2007} or the pyrochlore lattice\cite{Guo2009} in three dimensions, show flat surface bands with potential relevance for the iridate systems.\cite{Pesin2010}
As discussed in the following, a certain chiral symmetry in these models guarantees the presence of exactly flat surface bands.\cite{Volovik2009a,Heikkila2011b}

\begin{figure*}[tb]
\centering
\includegraphics[width=0.45\textwidth]{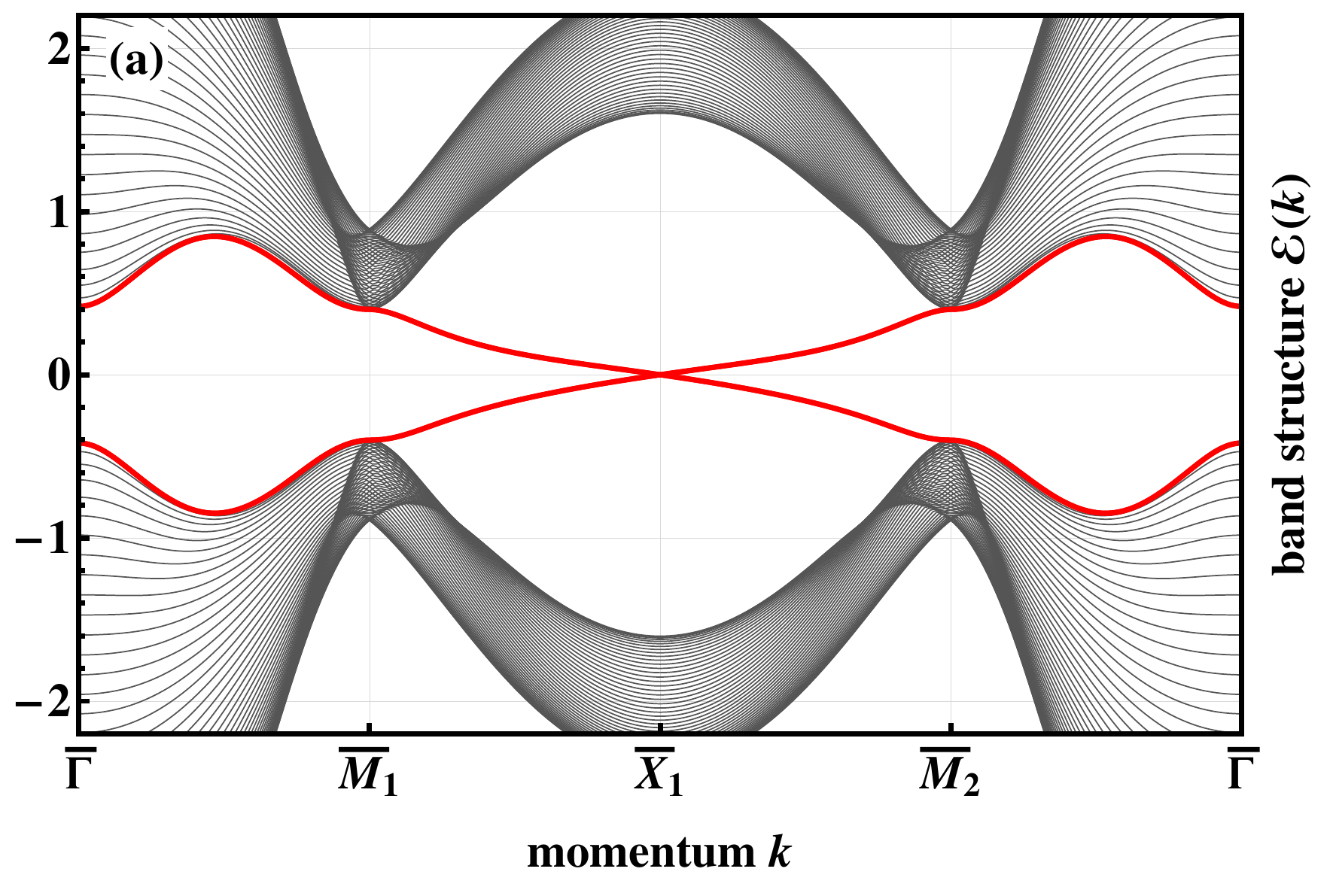}
\hspace*{0.05\textwidth}
\includegraphics[width=0.45\textwidth]{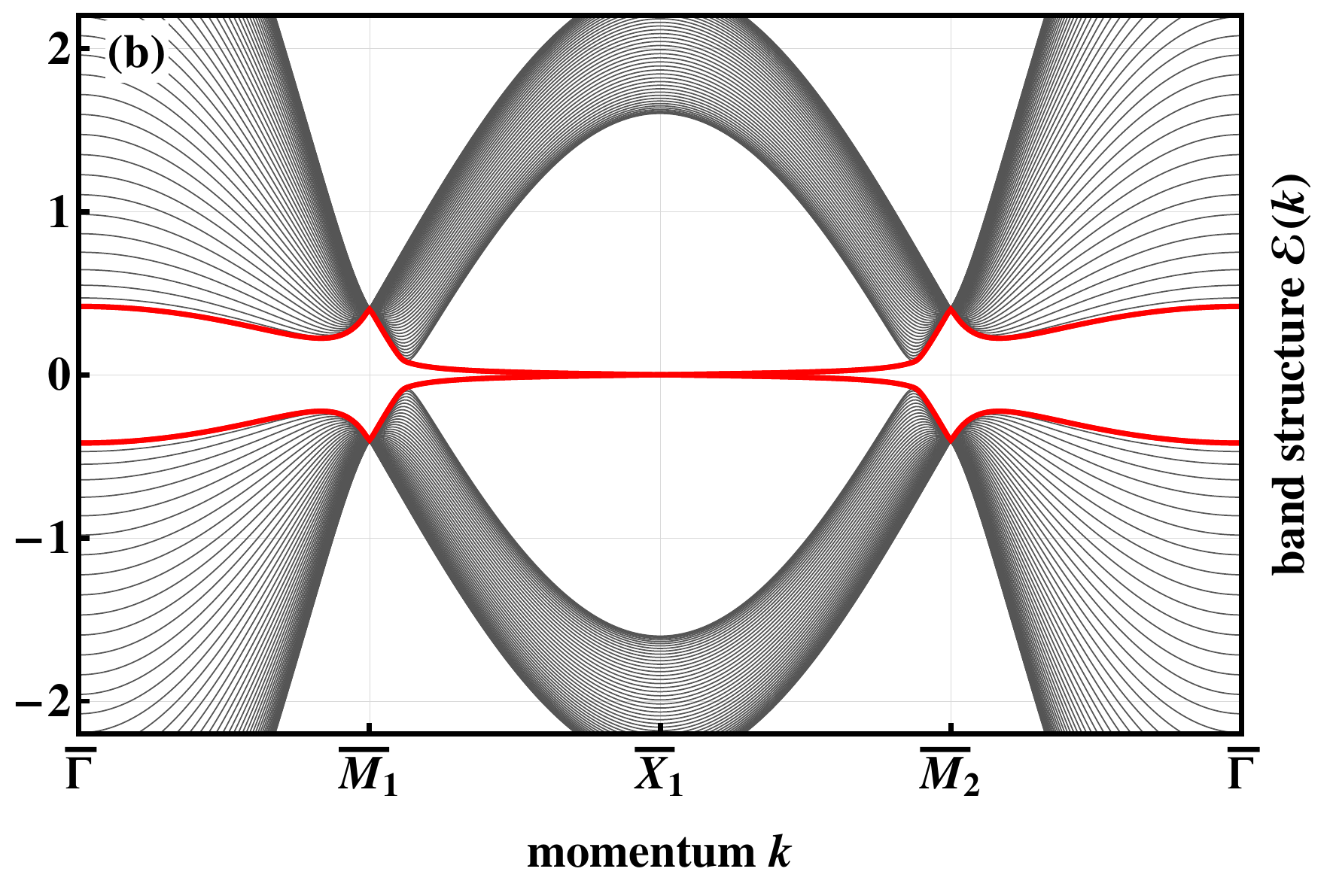}
\caption{%
(Color online)
Band structure of model~II [Eq.~(\ref{eq:FuKaneMeleModel})] for a slab with $(001)$ oriented surfaces with $t = 1$, $\delta t_{1}/t = 0.4$, and \textbf{(a)} $\lambda_{\mathrm{SO}}/t = 0.1$ and \textbf{(b)} $\lambda_{\mathrm{SO}}/t = 0.01$.
Upon decreasing the spin-orbit coupling $\lambda_{\mathrm{SO}}/t \to 0$, the surface states become flatter and flatter.
}
\label{fig:FuKaneMeleModel}
\end{figure*}

Our model~II, introduced by Fu, Kane, and Mele,\cite{Kane2005a, Fu2007} is defined on a diamond lattice
\begin{align}
\label{eq:FuKaneMeleModel}
\nonumber
H &= -\sum_{\langle i j \rangle, \sigma} (t_{ij} c_{i \sigma}^{\dagger} c_{j \sigma}^{\pdagger} + \mathrm{H.c.})  \\
&\quad + i \lambda_{\mathrm{SO}} \sum_{\langle\langle i j \rangle\rangle, \sigma, \sigma'} c_{i \sigma}^{\dagger} (\vect{\nu}_{ij}^{\pdagger} \cdot \vect{\sigma}_{\sigma \sigma'}^{\pdagger}) c_{j \sigma'}^{\pdagger}.
\end{align}
Here, $t_{ij} = t + \delta t_{\alpha}$ denotes the overlap parameter for hopping between nearest neighbors, where $\delta t_{\alpha}$ describes changes of the hopping parameter along the four nearest-neighbor bonds, $\alpha = 1, \dotsc, 4$, of the diamond lattice corresponding to distortions of the diamond lattice.
The second term is the spin-orbit interaction connecting second-nearest neighbors with a spin-dependent amplitude given by $\vect{\nu}_{ij} \equiv (\vect{d}_{ij}^{(1)} \times \vect{d}_{ij}^{(2)})/(a^{2}/8)$, where $\vect{d}_{ij}^{(1,2)}$ are the two (normalized) nearest-neighbor bond vectors traversed between sites $i$ and $j$.
The normalization of $\vect{\nu}_{ij}$ ensures that each component of the vector $\vect{\nu}_{ij}$ is of unit magnitude, \textit{i.e.}, $\nu_{ij}^{x,y,z} = \pm 1$.
After Fourier transformation, the Hamiltonian~(\ref{eq:FuKaneMeleModel}) takes the form
\begin{equation}
\label{eq:FuKaneMeleModel2}
\mcH(\vect{k}) = \vect{d}(\vect{k}) \cdot \vect{\Gamma},
\end{equation}
where
\begin{equation}
\begin{gathered}
\vect{d}(\vect{k}) \equiv \bigl( \RE \gamma(\vect{k}), -\IM \gamma(\vect{k}), u_{x}(\vect{k}), u_{y}(\vect{k}), u_{z}(\vect{k}) \bigr)^{T}, \\
\vect{\Gamma} \equiv (\tau_{x} \otimes \sigma_{0}, \tau_{y} \otimes \sigma_{0}, \tau_{z} \otimes \sigma_{x}, \tau_{z} \otimes \sigma_{y}, \tau_{z} \otimes \sigma_{z})^{T}.
\end{gathered}
\end{equation}
Hopping along the bonds $\vect{d}_{\alpha}$ ($\alpha = 1, \dotsc, 4$) of the diamond lattice is described by the tight-binding function
\begin{equation}
\gamma(\vect{k}) = -\sum_{\alpha=1}^{4} (t + \delta t_{\alpha}) e^{i \vect{k} \cdot \vect{d}_{\alpha}},
\end{equation}
while
\begin{equation}
u_{x}(\vect{k}) = 4 \lambda_{\mathrm{SO}} \sin(k_{x}/2) \bigl( \cos(k_{y}/a) - \cos(k_{z}/2) \bigr)
\end{equation}
describes the mixing of spin-up and spin-down states due to spin-orbit interactions; $u_{y}(\vect{k})$ and $u_{z}(\vect{k})$ are obtained by cyclic permutation of the lower indices.
As the components of $\vect{\Gamma}$ anti-commute, $\{ \Gamma_{a}, \Gamma_{b} \} = 2 \delta_{ab}\ \openone$, the spectrum of the Hamiltonian~(\ref{eq:FuKaneMeleModel2}) is easily obtained as
\begin{equation}
\mcE_{\pm}(\vect{k}) = \pm |\vect{d}(\vect{k})| = \pm \sqrt{\sum_{j=1}^{5} d_{j}(\vect{k})^{2}},
\end{equation}
where each band is doubly degenerate.

To obtain an insulator, the spin-orbit coupling parameter $\lambda_{\mathrm{SO}}$ has to be finite and there has to be a finite anisotropy: we choose $\delta t_{1}/t = 0.4$ and $\delta t_{2,3,4}/t = 0$ as in Ref.~[\onlinecite{Fu2007}].
For $\delta t_{1} > 0$ and $\lambda_{\mathrm{SO}} \not= 0$, one gets a strong TI.
For $\delta t_{\alpha} = 0$ and $\lambda_{\mathrm{SO}} \not= 0$, there are bulk Dirac points at the $\Gamma$ and the three $X$ points, $X_{\alpha} = \frac{2\pi}{a} \vect{e}_{\alpha}$ with $\alpha = x, y, z$.
For $\lambda_{\mathrm{SO}} = 0$ and $\delta t_{1} > 0$, the model is characterized by a special chiral symmetry discussed below which leads to lines in momentum space, where the gap vanishes.

\subsection{Topological origin and flatness of surface bands}

We have implemented the above Hamiltonian for two slab geometries with $(001)$ and $(111)$ surfaces.
For both slab orientations we observe that the surface states become increasingly flat as we decrease the spin-orbit coupling $\lambda_{\mathrm{SO}}/t$.
Both the bulk gap and the surface velocities scale linear in $\lambda_{\mathrm{SO}}$ (see Fig.~\ref{fig:FuKaneMeleModel}).

The existence of the flat surface bands is rooted in an additional ``chiral symmetry'' present at $\lambda_{\mathrm{SO}}/t = 0$ which is discussed in detail in Appendix~\ref{app:AdditionalChiralSymmetry}.
This symmetry leads to a bulk nodal line for $\lambda_{\mathrm{SO}}/t = 0$.
Furthermore, this chiral symmetry allows to define a winding number $\nu(\vect{k}_{\|})$ for each momentum $\vect{k}_{\|}$ of the surface Brillouin zone.\cite{Volovik2009a,Heikkila2011b}
For all surface momenta with $\nu(\vect{k}_{\|}) \not= 0$ there exists a zero-energy surface state.
The projection of the bulk nodal lines onto the surface define the boundaries of regions with finite $\nu(\vect{k}_{\|})$ (see Appendix~\ref{app:AdditionalChiralSymmetry}).
Generically, one obtains therefore a region in momentum space with exactly flat surface bands.
Note that this chiral symmetry has also been used previously in a closely related system predicting the presence of flat edge states in graphene ribbons with zigzag edges\cite{Ryu2002}.
More generally, any system with (i) only nearest-neighbor hopping of electrons on a bipartite lattice and (ii) a chiral symmetry $\hat{\Sigma}$ (as described in Appendix~\ref{app:AdditionalChiralSymmetry}), has topologically protected zero-energy boundary states.\cite{Gurarie2011}

Both spin-orbit interactions or a hopping $t'$ between second-nearest neighbors break, however, the additional chiral symmetry.
Consequently, the surface bands acquire a finite kinetic energy which we find numerically to vary linearly with $\lambda_{\mathrm{SO}}$ (see Fig.~\ref{fig:FuKaneMeleModel}) and $t'$ (not shown).
Therefore, the Fermi velocity at the surface can be estimated as
\begin{equation}
\frac{v_{F}^{\mathrm{surf}}}{t \, a} \sim  \max\left(\frac{t'}{t} ,\frac{\lambda_{\mathrm{SO}}}{t}\right) \ll 1.
\end{equation}
Generically, the surface velocity is therefore finite, but small.
Typically, both second-nearest-neighbor hopping and spin-orbit interactions are small.

This situation is reminiscent of the flat bands in particle-hole-symmetric graphene with zigzag edges.
Also in this case, second-nearest-neighbor hopping $t'$ breaks particle-hole symmetry and renders the surface state dispersive with a bandwidth set by $t'$ (Ref.~\citenum{Neto2009}).
This happens here as well, preventing the surface states from becoming perfectly flat when either $\lambda_{\mathrm{SO}}/t \not= 0$ or $t'/t \not= 0$.
Note that a finite $t'$ can not induce a TI for $\lambda_{\mathrm{SO}}/t = 0$.
Therefore, we focus, for simplicity, the following discussion on the limit $t'/t = 0$ and comment only briefly on the role of $t'$.

\subsection{Long-range Coulomb interactions and screening}

While we find surface bands with a small Fermi velocity, this does not automatically imply that $\alpha$ becomes large.
As for model I, screening effects become important.
We therefore calculate the polarization function (see Appendix~\ref{app:PolarizationFunctionModelII}) for small, but finite $\lambda_{\mathrm{SO}}/t$ and $t'/t = 0$ based on an expansion around the nodal lines obtained for $\lambda_{\mathrm{SO}}/t = 0$
We obtain
\begin{equation}
\Pi(\omega = 0, \vect{q}) \propto \frac{|\vect{q}|^{2}}{\sqrt{|\vect{q}|^{2} + \lambda_{\mathrm{SO}}^{2}}}\;.
\end{equation}
According to Eq.~(\ref{eps}), the dielectric constant, $\epsilon \sim 1 /\lambda_{\mathrm{SO}}$, is therefore of order $1/\lambda_{\mathrm{SO}}$.
Consequently, the leading $\lambda_{\mathrm{SO}}$ dependence of the Fermi velocity at the surface and the dielectric constant cancel for the effective fine-structure constant
\begin{equation}
\alpha = \frac{1}{\hbar \epsilon_{0} \epsilon v_{F}^{\mathrm{surf}}} \sim 1 \quad {\mathrm{as}} \quad \lambda_{\mathrm{SO}} \to 0.
\end{equation}
The value of $\alpha$ depends on microscopic details but remains of order $1$ even in the limit when the surface bands become flat.

\subsection{Effect of local interactions}

While $\alpha$ remains finite, this does not imply that a similar statement holds for short-ranged interactions.
In contrast to model $I$, where the surface bands cease to exist when the bulk gap closes, flat surface bands are enforced by topology for $\lambda_{\mathrm{SO}} = 0$.
Therefore, we expect that the penetration depth of surface states remains finite for large portions of the surface Brillouin zone in the limit $\lambda_{\mathrm{SO}} \to 0$ when the surface bands become flat (as we checked numerically).
In contrast to model I, the effective surface interaction therefore remains finite in the flat-band limit.
This also implies that
\begin{equation}
\label{eq:SurfaceInteraction}
\frac{U_{\mathrm{surf}}}{D_{\mathrm{surf}}} \sim \frac{U}{\lambda_{\mathrm{SO}}}.
\end{equation}
Thus, the critical interaction strength $U$ for surface magnetism is expected to vary linearly with $\lambda_{\mathrm{SO}}$.

To confirm this picture, we model the local interactions by an on-site Hubbard interaction
\begin{equation}
H_{\mathrm{int}} = U \sum_{i} \hat{n}_{i \uparrow}^{\pdagger} \hat{n}_{i  \downarrow}^{\pdagger},
\end{equation}
and calculate, as above, the layer-resolved susceptibility matrix for our slab geometries:
\begin{equation}
\label{eq:SusceptibilityMatrix2}
\chi_{j,j'}(\vect{q}) = \int \frac{d^{2}k}{(2\pi)^{2}} \!\!\! \sum_{\mcE_{\vect{k},\alpha} < 0 < \mcE_{\vect{k} + \vect{q}, \beta}} \!\!\!\!\!\!\!\! \frac{M^{\alpha \beta}_{\vect k,\vect k+\vect q}(j)M^{\beta \alpha}_{\vect k+\vect q,\vect k}(j')}{\mcE_{\vect{k}, \alpha} - \mcE_{\vect{k} + \vect{q}, \beta}}.
\end{equation}
Here, $M^{\alpha \beta}_{\vect{k}, \vect{k}'}(j) = \langle{\vect{k} \alpha}| \sigma_z(j) |{\vect{k}' \beta}\rangle$ denotes the matrix elements of the spin operator in layer $j$.
Comparing the eigenvalues of $\chi$ with $1/U$ allows us to extract the phase diagram as discussed above.

\begin{figure}[tb]
\centering
\includegraphics[width=0.45\textwidth]{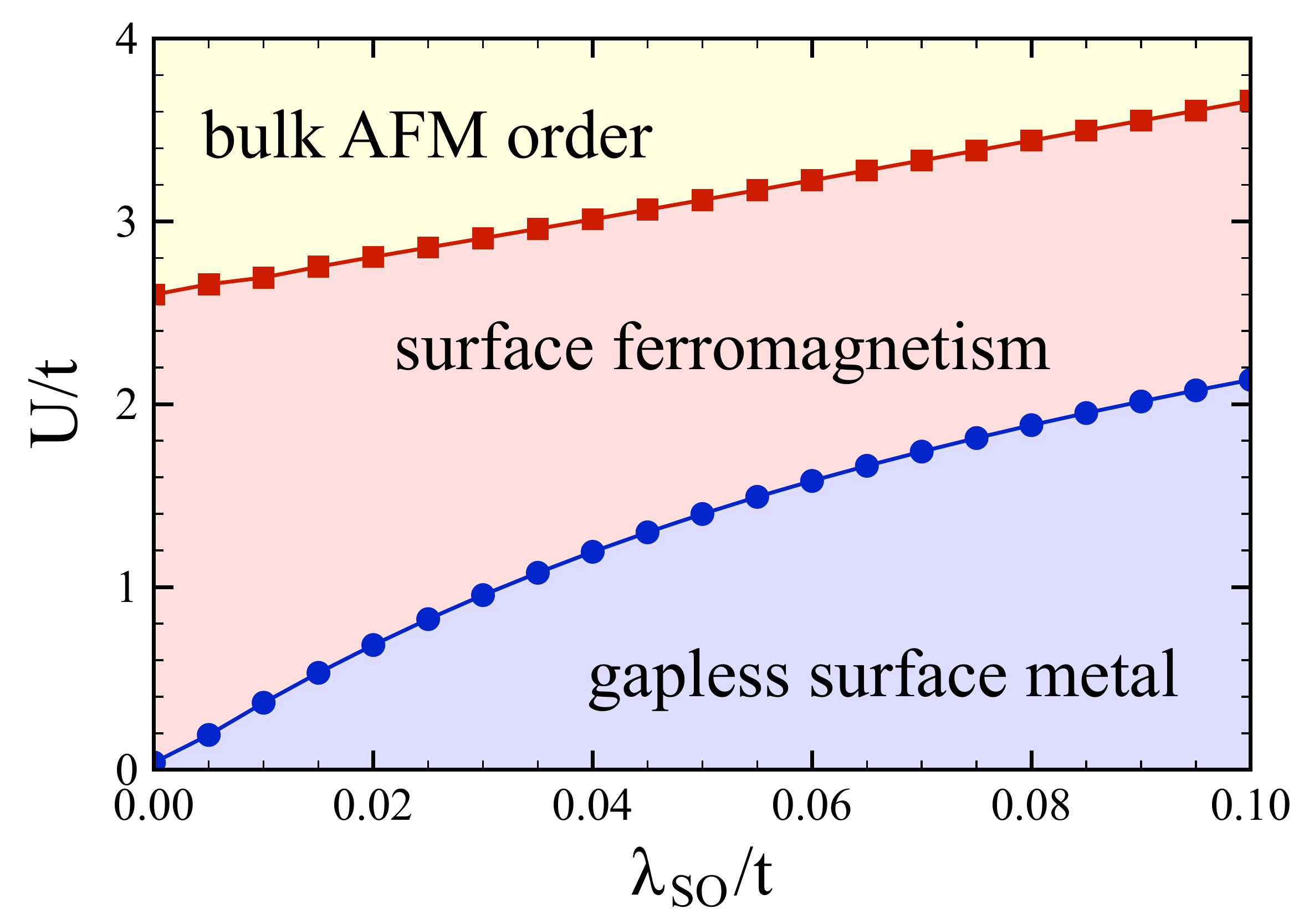}
\caption{%
(Color online)
Qualitative phase diagram of model~II, computed for a finite slab with $(001)$ surfaces and $N_{z} = 40$ epilayers for $t = 1$ and $\delta t_{1}/t = 0.4$ as a function of the interaction strength $U$ and the spin-orbit coupling constant $\lambda_{\mathrm{SO}}$.
A large part of the phase diagram is dominated by surface ferromagnetism which induces an anomalous quantum Hall effect.
While the critical interaction strength for surface ferromagnetism $U_{c}^{\mathrm{surf}} \sim v_{F}^{\mathrm{surf}} \sim \lambda_{\mathrm{SO}}$ goes to zero upon approaching the bulk QCP $\lambda_{\mathrm{SO}}/t = 0$ (blue circles), the critical $U_{c}^{\mathrm{bulk}}$ for bulk anti-ferromagnetism remains finite (red squares).
The transition line to bulk anti-ferromagnetic order has been calculated from a bulk mean field theory, but up to small finite-size effects the same result is obtained from the slab calculation.
}
\label{fig:PhaseDiagramFkmModel}
\end{figure}

The resulting qualitative phase diagram for model~II is shown in Fig.~\ref{fig:PhaseDiagramFkmModel}.
The most important observations are as follows:
(i)~In the limit $\lambda_{\mathrm{SO}} \to 0$ when the surface bands become flat, the critical value for surface ferromagnetism $U_{c}^{\mathrm{surf}}$ goes linearly to zero, while (ii) in the same limit the critical value $U_{c}^{\mathrm{bulk}}$ for bulk anti-ferromagnetism remains finite.
Therefore a large part of the phase diagram is dominated by surface ferromagnetism which induces an anomalous quantum Hall effect at the surface as described in the introduction.
This is consistent with the analytical arguments given above and in Eq.~(\ref{eq:SurfaceInteraction}).
In contrast to model~I, the surface bands in model~II remain confined to the surface in the limit when the surface velocity vanishes.
Therefore, surface magnetism naturally occurs in a wide parameter range.
While predictions within mean-field theory and RPA are quantitatively not reliable to predict magnetism, the qualitative statement that surface ferromagnetism occurs as the leading instability in distorted diamond lattices is expected to hold whenever particle-hole symmetry-breaking hoppings between second-nearest neighbors remain small.

\section{Conclusions}
\label{sec:Conclusions}

In this paper, we have investigated the question as to whether one can use a simple design principle to create TIs, where surface magnetism induces an anomalous quantum Hall effect.
At present, such materials are not known, but theory predicts many intriguing properties of such systems\cite{Qi2011, Qi2008, Essin2009}.

A comparison of  Figs.~\ref{fig:SusceptibilityMatrix} and \ref{fig:PhaseDiagramFkmModel} illustrates the main message of this paper:
In both cases, the Fermi velocity of surface bands of TIs is tuned to very small values and simultaneously the bulk gap closes.
In both cases, long-ranged Coulomb interactions remain sufficiently screened that, generically, they do not drive any phase transition.
Local interactions, however, stabilize in one case surface ferromagnetism in a large parameter range (Fig.~\ref{fig:PhaseDiagramFkmModel}), but not in the other case (Fig.~\ref{fig:SusceptibilityMatrix}).
This is explained by the observation that for the first example the wave functions of the surface states extend more and more into the bulk when the surface velocity goes to zero, while this is not the case in the second example.
One can therefore conclude that materials with an approximate chiral symmetry which provides flat surface bands (the case considered in Fig.~\ref{fig:PhaseDiagramFkmModel}) are ideal candidates to search for TIs with strongly interacting surfaces.

While we focused our study on the phase diagram at zero temperatures, similar results are expected to hold for finite temperatures.
Due to spin-orbit coupling and the resulting Ising symmetry of surface magnetism, thermal fluctuations do not give rise to any singular corrections.
Therefore, also as a function of temperature, the leading instability of model~II is expected to be surface ferromagnetism.

The most promising alternative route to surface magnetism is to use chemistry: doping the surface with magnetic elements or growing monolayers of magnetic films on the surfaces of TIs will likely be able to induce surface magnetism.
To stabilize more exotic phases (\textit{e.g.}, fractional Hall phases), however, reducing the kinetic energy on the surface without inducing too much screening seems to be desirable.
While we have restricted our investigation to magnetism, one can expect that our main conclusions will also hold for other exotic phases at the surface.

\section*{Acknowledgments}

We acknowledge discussions with S.~Ryu, A.~Schnyder, and B.~Trauzettel.
We were supported by the DFG within FOR 960, SFB TR12 (AR, MS), BCGS (MS), and FR 2627/3-1 (LF).

\appendix

\section{Dielectric constant and polarization function in the random phase approximation}

\begin{figure}[tb]
\centering
\includegraphics[width=0.33\textwidth]{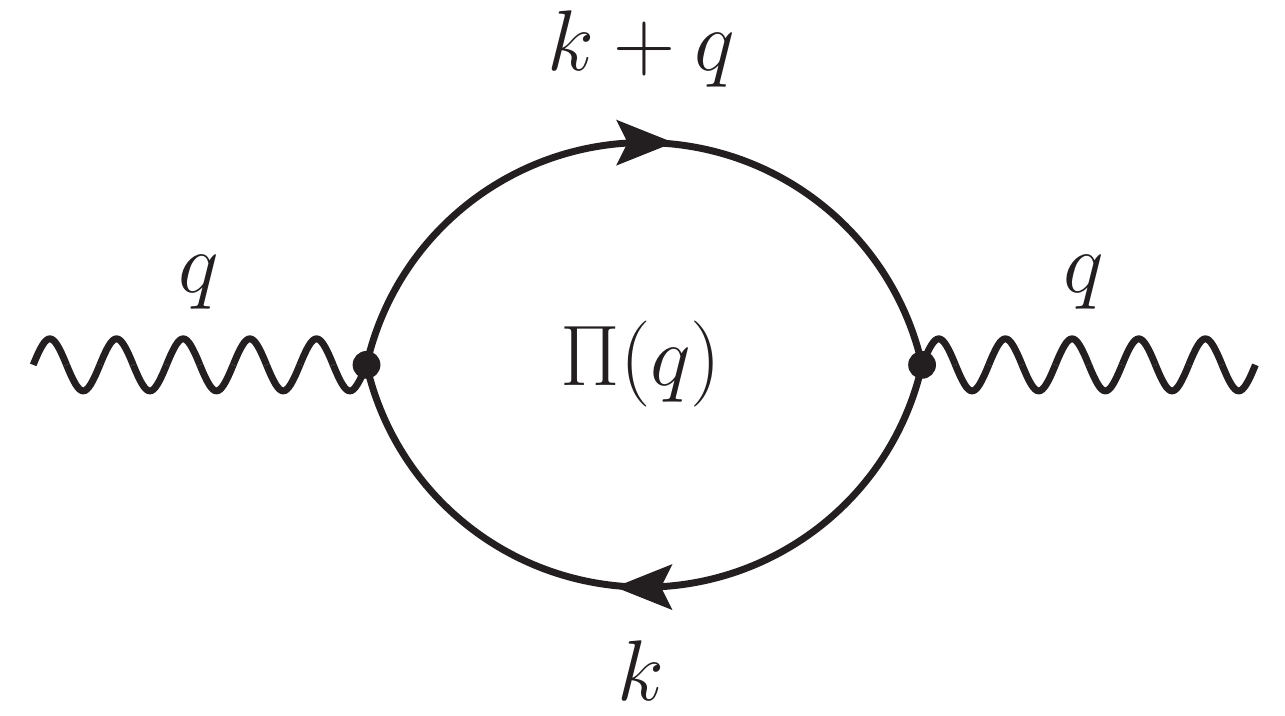}
\caption{%
Feynman diagram for the polarization function $\Pi(q) = \Pi(\omega, \vect{q})$.
The ``4-momenta'' $k$ and $q$ comprise both frequency and momentum, \textit{i.e.}, $k \equiv (\nu, \vect{k})$ and $q \equiv (\omega, \vect{q})$.
}
\label{fig:PolarizationBubble}
\end{figure}

Within the random phase approximation, the dielectric constant $\epsilon$ is related to the polarization function $\Pi$ by the relation
\begin{equation}
\epsilon(\omega, \vect{q}) = 1 + \frac{4\pi e}{|\vect{q}|^{2}}\ \Pi(\omega, \vect{q}).
\end{equation}
In Matsubara frequencies and at zero temperature $\Pi$ can be rewritten as a four-dimensional momentum integral (see Fig.~\ref{fig:PolarizationBubble}):
\begin{equation}
\Pi(\omega, \vect{q}) = 2 \int \frac{d\nu}{2\pi} \int \frac{d^{3}k}{(2\pi)^{3}} \trace \{ \mcG_{0}(\nu, \vect{k}) \mcG_{0}(\nu + \omega, \vect{k} + \vect{q}) \},
\end{equation}

To determine the screening properties of both models I and II close to the bulk quantum critical point, we calculate in the following the polarization function $\Pi$ in the static limit and for long wavelengths, $\lim_{\vect{q}\to 0}\Pi(\omega = 0, \vect{q})$.

\subsection{Model~I}\label{app:PolarizationFunctionModelI}

First, let us discuss the polarization function for model~I close to the bulk quantum critical point $m/t = 3$ in the presence of inversion asymmetry, \textit{i.e.}, $\Delta/t \not= 0$.
For $\delta m/t \equiv |m - 3t|/t \ll 1$, we may consider a linearized version of the minimal model:
\begin{equation}
\mcH_{\mathrm{eff}}(\vect{k}) \approx \delta m\ \Gamma_{0} - a t \sum_{j=1}^{3} k_{j}\ \Gamma_{j} + \Delta\ \Gamma_{04},
\end{equation}
where $\vect{k}$ measures the distance to the $\Gamma$~point, \textit{i.e.}, the center of the Brillouin zone.
The free Green's function is defined as the single-particle propagator
\begin{equation}
\mcG_{0}(k) \equiv \langle \Psi_{k}^{\dagger} \Psi_{k}^{\pdagger} \rangle_{0} = \bigl( -i \nu\, \openone + \mcH_{\mathrm{eff}}(\vect{k}) \bigr)^{-1},
\end{equation}
where $k \equiv (\nu, \vect{k})$ is the 4-momentum consisting of both frequency $\nu$ and momentum $\vect{k}$.
Let $\mcU(\vect{k})$ be the unitary matrix of normalized eigenvectors of the effective Hamiltonian $\mcH_{\mathrm{eff}}(\vect{k})$ such that
\begin{equation}
\mcU(\vect{k}) \mcH_{\mathrm{eff}}(\vect{k}) \mcU(\vect{k})^{\dagger} = \diag(\{ \mcE_{n}(\vect{k}) \}).
\end{equation}
Here, $\mcE_{n}(\vect{k})$ denotes the eigenenergies of $\mcH_{\mathrm{eff}}(\vect{k})$:
\begin{equation}
\label{eq:SQC:LinearizedBands}
\mcE_{\pm \pm}(\vect{k}) \equiv \pm \sqrt{\delta m^{2} + (a t |\vect{k}| \pm \Delta)^{2}}.
\end{equation}
As a consequence, the free Green's function $\mcG_{0}(k)$ can be rewritten as
\begin{subequations}
\begin{align}
\mcG_{0}(k) &= \mcU(\vect{k})^{\dagger}  \mcD(k) \mcU(\vect{k}) , \\
\mcD(k) &\equiv \diag \bigl( \{ (-i \nu + \mcE_{n}(\vect{k}))^{-1} \} \bigr).
\end{align}
\end{subequations}
Upon substituting $\mcG_{0}(k)$ we obtain the following expression for the polarization function $\Pi$:
\begin{multline}
\Pi(\omega, \vect{q}) = 2 \int \frac{d\nu}{2\pi} \int \frac{d^{3}k}{(2\pi)^{3}} \\
\trace \{ \mcF(\vect{k}, \vect{q})^{\dagger} \mcD(k) \mcF(\vect{k}, \vect{q}) \mcD(k+q) \},
\end{multline}
where, for brevity, we have introduced the ``structure factor'' $\mcF(\vect{k}, \vect{q}) \equiv \mcU(\vect{k})^{\dagger} \mcU(\vect{k} + \vect{q})$ which depends on the momenta $\vect{k}$ and $\vect{q}$.
Since $\mcD(k)$ is diagonal, the polarization function $\Pi$ then simply reads
\begin{multline}
\Pi(\omega, \vect{q}) = 2 \int \frac{d\nu}{2\pi} \frac{d^{3}k}{(2\pi)^{3}} \sum_{m,n=1}^{4} \\
\frac{[\mcF(\vect{k}, \vect{q})^{\dagger}]_{mn} \mcF(\vect{k}, \vect{q})_{nm}}{(i \nu - \mcE_{n}(\vect{k})) (i \nu + i \omega - \mcE_{m}(\vect{k} + \vect{q}))}.
\end{multline}
In the static limit ($\omega = 0$) and at zero temperature, the integration over the internal frequency $-\infty < \nu < \infty$ can be easily computed using the residual theorem:
\begin{multline}
\Pi(\omega=0, \vect{q}) = 2 \int \frac{d^{3}k}{(2\pi)^{3}} \sum_{m,n=1}^{4} [\mcF(\vect{k}, \vect{q})^{\dagger}]_{mn} \mcF(\vect{k}, \vect{q})_{nm} \\
\times \frac{\Theta(-\mcE_{m}(\vect{k} + \vect{q})) - \Theta(-\mcE_{n}(\vect{k}))}{\mcE_{m}(\vect{k} + \vect{q}) - \mcE_{n}(\vect{k})},
\end{multline}
where $\Theta(x)$ denotes the Heaviside theta function.
Consequently, the only non-vanishing contributions to the static polarization function in the gapped system stem from particle-hole excitations:
\begin{multline}
\Theta(-\mcE_{m}(\vect{k} + \vect{q})) - \Theta(-\mcE_{n}(\vect{k})) \\
= \begin{cases}
    1 & \text{$m$ occupied and $n$ empty} \\
    -1 & \text{$m$ empty and $n$ occupied} \\
    0 & \text{both $m$ and $n$ empty or occupied}
  \end{cases}
\end{multline}
Note that the energy denominator $\mcE_{m}(\vect{k} + \vect{q}) - \mcE_{n}(\vect{k})$ is non-zero as long as $\delta m \not= 0$, because the indices $m$ and $n$ refer to either valence or conduction bands, respectively.

Concerning the quantum critical point $\delta m = 0$ of model~I, it turns out that the static polarization function $\Pi(\omega=0, \vect{q} \to 0)$ develops a singularity in the long-wavelength limit which is a precursor to screening of long-range Coulomb interactions.
In the following, we consider the inversion asymmetry $\Delta\ \Gamma_{04}$ as a small perturbation to the strong TI, \textit{i.e.}, $\Delta/t \ll 1$, and focus on the leading-order terms of $\Pi$.
Starting from the linearized band structure~(\ref{eq:SQC:LinearizedBands}) we see that in the vicinity of the bulk Dirac point the singular contribution to the polarization function stems from the touching of the highest valence band and the lowest conduction band:
\begin{equation}
\mcE_{\pm,-}(\vect{k}) = \pm \sqrt{\delta m^{2} + (a t |\vect{k}| - \Delta)^{2}},
\end{equation}
where, for simplicity, we have assumed that $\Delta$ is positive.
The divergence of the polarization is due to the fact that in the limit $\vect{q} \to 0$ and $\delta m \to 0$ the energy denominator vanishes, while the other contributions to $\Pi$ remain largely constant or vanish quadratically for small $\vect{q}$ vectors, and therefore we neglect those terms in the following discussion.
The divergent term $\Pi^{(23)}(\omega=0, \vect{q})$ of the polarization function takes the form
\begin{equation}
\Pi^{(23)}(\omega=0, \vect{q}) = 2 \int \frac{d^{3}k}{(2\pi)^{3}} \frac{[\mcF(\vect{k}, \vect{q})^{\dagger}]_{23} \mcF(\vect{k}, \vect{q})_{32}}{\mcE_{2}(\vect{k} + \vect{q}) - \mcE_{3}(\vect{k})},
\end{equation}
where $\mcE_{2,3}(\vect{k}) \equiv \mcE_{\pm,-}(\vect{k})$ as introduced above.
This integral can be solved analytically, for example by introducing a suitable coordinate system such that $\vect{q} = q \vect{e}_{z}$ and spherical coordinates $\{ k, \theta, \phi \}$ for the momentum integral over $\vect{k}$.
After substituting $x = \cos \theta$ with corresponding integral measure $dx = -\sin \theta\ d\theta$ we perform a Taylor expansion of both numerator and denominator for small $q \ll \delta m/t \ll \Delta/t \ll 1$.
As a result, the relevant term takes the form
\begin{multline}
\Pi^{(23)}(\omega=0, \vect{q}) = \frac{1}{16 \pi^{2}}\ (a t)^{2} |\vect{q}|^{2} \\
\times \int_{0}^{\Lambda} dk\ k^{2} \int_{-1}^{1} dx\ \frac{x^{2}\ \delta m^{2}}{[(a t k - \Delta)^{2} + \delta m^{2}]^{5/2}}.
\end{multline}
Note that the integration over $x \in [-1, 1]$ simply gives a prefactor of $2/3$, and the remaining integral over $k \in [0, \infty)$ also can be easily performed, resulting in:
\begin{equation}
\Pi^{(23)}(\omega=0, \vect{q}) = \frac{1}{144 \pi^{2}}\ (a t)^{2} |\vect{q}|^{2} \biggl( \frac{\Delta + \sqrt{\delta m^{2} + {\Delta}^{2}}}{\delta m\ t^{3/2}} \biggr)^{2}.
\end{equation}
Close to the bulk quantum critical point, we may simplify our result for $\delta m/t \ll \Delta/t \ll 1$, such that we obtain for the full polarization function:
\begin{equation}
\begin{split}
\Pi(\omega=0, \vect{q}) &\approx \frac{1}{18 \pi^{2}}\ (a t)^{2} |\vect{q}|^{2} \frac{{\Delta}^{2}}{t^{3}\ \delta m^{2}} \\
&= \frac{1}{18 \pi^{2}} \frac{{\Delta}^{2}}{t\ \delta m^{2}}\ (v_{F}^{\mathrm{bulk}} |\vect{q}|)^{2},
\end{split}
\end{equation}
where $v_{F}^{\mathrm{bulk}} = t a$ is the bulk Fermi velocity.
We have also taken the factor of $2$ into account which arises from the fact that $\Pi(\omega=0, \vect{q})_{23} = \Pi(\omega=0, \vect{q})_{32}$.
Note that the prefactor $1/(18 \pi^{2})$ agrees with numerical calculations of the full polarization function $\Pi(\omega, \vect{q})$ in the low-energy limit.
The above result shows that the static polarization function $\Pi(\omega=0, \vect{q} \to 0)$ diverges quadratically as function of $\delta m$.
Consequently, this implies that the bulk dielectric constant $\epsilon$ diverges as well:
\begin{equation}
\epsilon \propto \mathrm{const.} + \frac{{\Delta}^{2}}{\delta m^{2}} \to \infty
\quad \text{as} \quad \delta m \to 0.
\end{equation}
Hence, chiral symmetry breaking due to long-range Coulomb interactions is avoided, because the effective interaction strength $\alpha$ vanishes:
\begin{equation}
\alpha = \frac{e^{2}}{\hbar v_{F}^{\mathrm{surf}} \epsilon} \propto \frac{\delta m}{\Delta} \to 0
\quad \text{as} \quad \delta m \to 0
\end{equation}
as $v_{F}^{\mathrm{surf}} \propto \delta m/\Delta$.
This particular behavior can be traced back to the nearby presence of the bulk states which leads to screening of the long-range Coulomb interaction.

\subsection{Model~II}
\label{app:PolarizationFunctionModelII}

To calculate the polarization function $\Pi$ for model~II, we start from the Hamiltonian~(\ref{eq:FuKaneMeleModel}) and Fourier transform to momentum space.
The corresponding Bloch Hamiltonian $\mcH_{\mathrm{FKM}}(\vect{k})$ can be written in terms of five $4 \times 4$ matrices $\Gamma_{a}$ ($a = 1, \dotsc, 5$) as
\begin{equation}
\mcH_{\mathrm{FKM}}(\vect{k}) = \vect{d}(\vect{k}) \cdot \vect{\Gamma},
\end{equation}
where the five-component vectors $\vect{d}(\vect{k})$ and $\vect{\Gamma}$ are defined as follows:
\begin{subequations}
\begin{align}
\vect{d}(\vect{k}) &\equiv \bigl( \RE \gamma(\vect{k}), -\IM \gamma(\vect{k}), \notag \\
&\qquad \lambda_{\mathrm{SO}} u_{x}(\vect{k}), \lambda_{\mathrm{SO}} u_{y}(\vect{k}), \lambda_{\mathrm{SO}} u_{z}(\vect{k}) \bigr)^{T}, \\
\vect{\Gamma} &\equiv (\tau_{x} \otimes \sigma_{0}, \tau_{y} \otimes \sigma_{0}, \tau_{z} \otimes \sigma_{x}, \tau_{z} \otimes \sigma_{y}, \tau_{z} \otimes \sigma_{z})^{T}.
\end{align}
\end{subequations}
Similar to model~I, those $\Gamma$ matrices satisfy the usual Clifford algebra, \textit{i.e.}\, they anti-commute as follows:
\begin{equation}
\{ \Gamma_{a}, \Gamma_{b} \} = 2 \delta_{ab}\ \openone
\quad \text{with} \quad
a, b = 1, \dotsc, 5.
\end{equation}
Due to the anti-com\-mu\-ta\-tiv\-i\-ty of the $\Gamma$ matrices, we obtain a particularly simple expression for the bare fermion propagator defined by:
\begin{equation}
\mcG_{0}(\nu, \vect{k}) \equiv \bigl( -i \nu\, \openone + \mcH(\vect{k}) \bigr)^{-1} = \frac{i \nu\, \openone + \vect{d}(\vect{k}) \cdot \vect{\Gamma}}{\nu^{2} + |\vect{d}(\vect{k})|^{2}}
\end{equation}
Furthermore, as the $\Gamma$ matrices are traceless ($\trace \Gamma_{a} = 0$ for $a = 1, \dotsc, 5$) we compute the matrix trace to obtain the following expression for the polarization function:
\begin{multline}
\Pi(\omega, \vect{q}) = 8 \int \frac{d\nu}{2\pi} \int \frac{d^{3}k}{(2\pi)^{3}} \\
\frac{-\nu (\nu + \omega) + \vect{d}(\vect{k}) \cdot \vect{d}(\vect{k} + \vect{q})}{[\nu^{2} + |\vect{d}(\vect{k})|^{2}] [(\nu + \omega)^{2} + |\vect{d}(\vect{k} + \vect{q})|^{2}]}.
\end{multline}
To compute the frequency integral over $-\infty < \nu < \infty$, we perform a rotation to the imaginary frequency axis by substituting $\nu = iz$.
Using the residue theorem, we then get for $\omega = 0$ the following exact result for the static polarization function:
\begin{multline}
\Pi(\omega=0, \vect{q}) = 4 \int \frac{d^{3}k}{(2\pi)^{3}} \\
\frac{|\vect{d}(\vect{k})| |\vect{d}(\vect{k} + \vect{q})| - \vect{d}(\vect{k}) \cdot \vect{d}(\vect{k} + \vect{q})}{|\vect{d}(\vect{k})| |\vect{d}(\vect{k} + \vect{q})| (|\vect{d}(\vect{k})| + |\vect{d}(\vect{k} + \vect{q})|)}.
\end{multline}
To compute the remaining momentum integral in three dimensions, we first observe that the above expression is well-defined as long as the system has a finite bulk band gap.
This is due to the fact that the electronic band structure of the model~II, $\mcH_{\mathrm{FKM}}(\vect{k}) = \vect{d}(\vect{k}) \cdot \vect{\Gamma}$, is given by $\mcE_{\pm}(\vect{k}) = \pm |\vect{d}(\vect{k})|$, where each band is doubly degenerate.
For finite spin-orbit interactions $\lambda_{\mathrm{SO}}/t$ and non-zero modulations of the hopping amplitudes $\delta t_{\alpha}/t$, the vector $\vect{d}(\vect{k})$ does not become singular and $|\vect{d}(\vect{k})|$ remains finite.
Consequently, we can expand the integrand for small wave vectors $\vect{q}$ so that the static polarization function takes the following form:
\begin{multline}
\Pi(\omega=0, \vect{q}) \approx 2 \int \frac{d^{3}k}{(2\pi)^{3}}\ q_{\alpha} q_{\beta}  \biggl[ \frac{(\partial_{\alpha} d_{\gamma}(\vect{k})) (\partial_{\beta} d_{\gamma}(\vect{k}))}{|\vect{d}(\vect{k})|^{3}} \\
- \frac{d_{\gamma}(\vect{k}) (\partial_{\alpha} d_{\gamma}(\vect{k})) d_{\delta}(\vect{k}) (\partial_{\beta} d_{\delta}(\vect{k}))}{|\vect{d}(\vect{k})|^{5}} \biggr] + \mcO(q^{3}),
\end{multline}
where $\partial_{\alpha} f(\vect{k}) \equiv \partial f(\vect{k})/\partial k_{\alpha}$, and a summation over the Greek indices $\alpha, \beta, \gamma, \delta \in \{ x, y, z \}$ is implied.
In a more compact notation:
\begin{multline}
\Pi(\omega=0, \vect{q}) \approx 2 \int \frac{d^{3}k}{(2\pi)^{3}} \biggl\{ \frac{[(\vect{q} \cdot \nabla_{\vect{k}}) \vect{d}(\vect{k})]^{2}}{|\vect{d}(\vect{k})|^{3}} \\
- \frac{[\vect{d}(\vect{k}) \cdot (\vect{q} \cdot \nabla_{\vect{k}}) \vect{d}(\vect{k})]^{2}}{|\vect{d}(\vect{k})|^{5}} \biggr\} + \mcO(q^{3}).
\end{multline}
This result for the static polarization function in the limit of long wave lengths, $\vect{q} \to 0$, is apparently well-defined as long as the bulk remains gapped which is ensured by a finite spin-orbit coupling $\lambda_{\mathrm{SO}}/t \not= 0$ and finite hopping modulations $\delta t_{\alpha}/t$, as explained above.
Furthermore, in the limit $\lambda_{\mathrm{SO}}/t \to 0$ nodal lines emerge in the bulk of the model~II close to the $X$ points, and those nodal lines also give rise to the perfectly flat surface states (cf.\ Appendix~\ref{app:AdditionalChiralSymmetry}).
Close to the bulk quantum critical point $\lambda_{\mathrm{SO}}/t = 0$, we can describe such a nodal line by an effective low-energy Hamiltonian
\begin{equation}
\mcH_{\mathrm{eff}}(\vect{k}) = \tilde{\vect{d}}(\vect{k}) \cdot \vect{\tau}
\quad \text{with} \quad{}
\tilde{\vect{d}}(\vect{k}) = (k_{x}, k_{y}, \lambda_{\mathrm{SO}})^{T},
\end{equation}
where we have performed a suitable coordinate transformation, so that $k_{z}$ is the momentum along the nodal line which does not enter the effective Hamiltonian.
Upon substituting the vector $\tilde{\vect{d}}(\vect{k})$ into the polarization function $\Pi(\omega=0, \vect{q})$ we find that
\begin{equation}
\Pi_{\mathrm{eff}}(\omega=0, \vect{q}) \propto \frac{q_{x}^{2} + q_{y}^{2}}{|\lambda_{\mathrm{SO}}|}.
\end{equation}
In the model~II, there are three equivalent $X$ points in the Brillouin zone of the diamond lattice, located at $X_{x} = \frac{2\pi}{a}\ (1,0,0)^{T}$, $X_{y} = \frac{2\pi}{a}\ (0,1,0)^{T}$, and $X_{z} = \frac{2\pi}{a}\ (0,0,1)^{T}$.
Hence, combining the above result for all three $X$ points we find for the static polarization function:
\begin{equation}
\Pi_{\mathrm{FKM}}(\omega=0, \vect{q}) \propto \frac{|\vect{q}|^{2}}{|\lambda_{\mathrm{SO}}|} + \mcO(q^{3}).
\end{equation}
This particular form of the polarization function is reminiscent of the situation in \textit{two}-dimensional graphene sheets with a single Dirac \textit{point} and mass $m$, where the static polarization function takes the following form:
\begin{equation}
\Pi_{\mathrm{graphene}}(\omega=0, \vect{q}) \propto \frac{|\vect{q}|^{2}}{\sqrt{v_{F}^{2} |\vect{q}|^{2} + m^2}}.
\end{equation}
Here, we find a similar relation in the \textit{three}-dimensional model~II for a \textit{line} of Dirac points, where the spin-orbit interaction generates a (sublattice-dependent) mass term for the otherwise gapless Dirac fermions:
\begin{equation}
\begin{split}
\Pi(\omega=0, \vect{q}) &\propto \frac{|\vect{q}|^{2}}{\sqrt{v_{F}^{2} |\vect{q}|^{2} + \lambda_{\mathrm{SO}}^2}} \\
&\propto
  \begin{cases}
    |\vect{q}|^{2}/|\lambda_{\mathrm{SO}}| & \text{for $\lambda_{\mathrm{SO}} \not=0$} \\
    |\vect{q}|/v_{F} & \text{for $\lambda_{\mathrm{SO}} = 0$}
  \end{cases}
\end{split}
\end{equation}
An important consequence of this result is that the dielectric constant $\epsilon$ diverges upon approaching the bulk quantum critical point $\lambda_{\mathrm{SO}}/t = 0$, which can be interpreted a precursor to the screening of long-range Coulomb interactions by the bulk nodal lines as $\lambda_{\mathrm{SO}} \to 0$:
\begin{equation}
\epsilon = 1 + \frac{4\pi e}{|\vect{q}|^{2}}\ \Pi(\omega=0, \vect{q}) \propto c_{0} + \frac{c_{1}}{|\lambda_{\mathrm{SO}}|} \to \infty
\end{equation}
with some numerical constants $c_{0}$ and $c_{1}$.
Furthermore, this also has a profound consequence on the effective interaction strength $\alpha$ describing long-range Coulomb interactions:
\begin{equation}
\alpha = \frac{e^{2}}{\hbar v_{F}^{\mathrm{surf}} \epsilon} \propto \frac{e^{2}}{\hbar} \times \frac{1}{|\lambda_{\mathrm{SO}}|} \times \frac{1}{c_{0} + c_{1}/|\lambda_{\mathrm{SO}}|} \propto \frac{1}{c_{1}},
\end{equation}
where we have used that the surface Fermi velocity close to the bulk critical point is roughly given by $v_{F}^{\mathrm{surf}} \propto \max \{ |\lambda_{\mathrm{SO}}|, t' \}$.
Notably, the effective interaction strength does not generically become large when $v_{F}^{\mathrm{surf}} \to 0$ upon approaching the quantum critical point, in stark contrast to the naive expectation that a flat surface band leads to a diverging interaction strength $\alpha$, thereby opening a surface band gap.
Hence, in general we do not expect a spontaneous mass generation due to long-range Coulomb interactions in model~II and similar models due to the screening of the long-range Coulomb interactions.

\section{Additional chiral symmetry in model~II}
\label{app:AdditionalChiralSymmetry}

\begin{figure*}[tb]
\centering
\includegraphics[height=5cm]{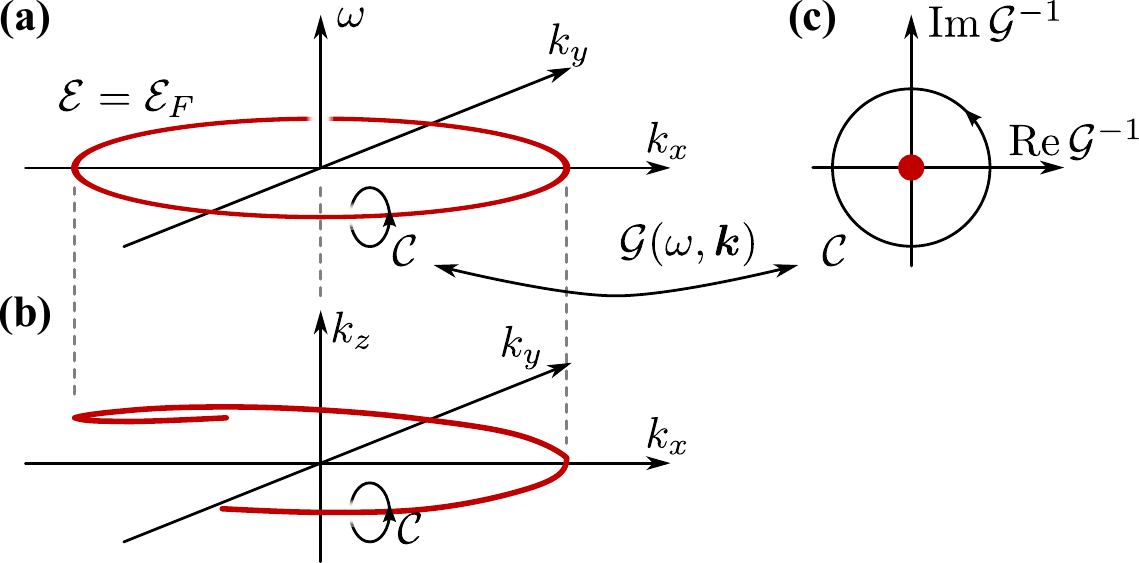}
\hspace*{0.05\textwidth}
\includegraphics[height=5cm]{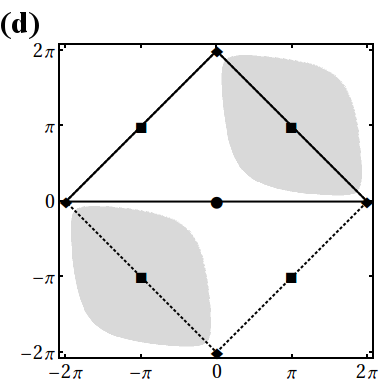}
\caption{%
\textbf{(a)--(c)}~Fermi surface $\mcE(\vect{k}) = \mcE_{F}$ (red line) as the momentum-space analog of a vortex line, where the phase of the zero-frequency Green's function changes by an integer multiple of $2\pi$ going around the nodal line in $(\omega, \vect{k})$ space.\cite{Volovik2009a}
The nodal line gives rise to topologically protected states when projected onto the surface Brillouin zone of a slab geometry.
\textbf{(d)}~Plot of the winding number $\nu(\vect{k}_{\|})$ as a function of $\vect{k}_{\|} = (k_{x}, k_{y})$ in the surface Brillouin zone for a $(001)$ oriented surface with $t = 1$ and $\delta t_{1}/t = 0.4$.
The surface Brillouin zone is indicated by dotted lines, while paths through the Brillouin zone are are marked by solid lines.
The surface points of high symmetry ($\overline{\Gamma}$, $\overline{X}$, and $\overline{M}$) are indicated by a circle, squares, and diamond symbols.
}
\label{fig:WindingNumber}
\end{figure*}
The topological invariant supporting the existence and topological stability of the flat surface bands in model~II is the following contour integral\cite{Volovik2009a, Heikkila2011b}:
\begin{equation}
\nu = \frac{1}{N} \oint_{\mcC} \frac{d\vect{k}}{2\pi i} \cdot \trace \bigl\{ \hat{\Sigma} \mcH(\vect{k})^{-1} \nabla_{\vect{k}} \mcH(\vect{k}) \bigr\} \in \mathbb{Z}\;,
\end{equation}
where $N$ denotes the number of conduction and valence bands, and $\hat{\Sigma}$ defines the chiral symmetry operator.
For model~II on a bipartite lattice we find $\hat{\Sigma} \equiv \tau_{z} \otimes \sigma_{0}$, where $\tau_{\alpha}$ and $\sigma_{\alpha}$ act on the sublattice and spin degrees of freedom, respectively.
The topological protection stems from the fact that the phase of the zero-energy Green's function $\mcG(\omega=0, \vect{k}) = \mcH(\vect{k})^{-1}$ can only change by an integer multiple of $2\pi$ when going around the bulk nodal line which is present only when $\lambda_{\mathrm{SO}}/t = 0$ and $t'/t = 0$, where $t'$ denotes hopping between second-nearest neighbors.

In a slab geometry, one may deform the integration contour $\mcC$ to a straight line $[-\pi/a, \pi/a]$ along the direction normal to the slab surface, and consider the conserved in-plane momentum $\vect{k}_{\|}$ in the surface Brillouin zone as a parameter to the winding number\cite{Heikkila2011b}:
\begin{equation}
\nu(\vect{k}_{\|}) = \frac{1}{N} \int_{-\pi/a}^{\pi/a} \frac{dk_{\perp}}{2\pi i} \trace \bigl\{ \hat{\Sigma} \mcH(k_{\perp}, \vect{k}_{\|})^{-1} \partial_{k_{\perp}} \mcH(k_{\perp}, \vect{k}_{\|}) \bigr\}.
\end{equation}
If we assume that for any point $\vect{k}_{\|}$ in the surface Brillouin zone within the projection of the nodal line $\nu(\vect{k}_{\|}) \not= 0$ and $\nu(\vect{k}_{\|}) = 0$ outside [see Fig.~\ref{fig:WindingNumber}~\textbf{(a)}], a zero-energy boundary state emerges for all momenta inside the non-trivial region since it cannot be connected adiabatically to the vacuum with trivial winding number $\nu = 0$.
The non-trivial winding number gives rise to large regions in the surface Brillouin zone, where the surface bands in model~II become perfectly flat [see Fig.~\ref{fig:WindingNumber}~\textbf{(d)}].

\end{document}